\documentclass[aps,twocolumn,floats,prd,nofootinbib,10pt]{revtex4-1}

\usepackage{graphicx,amsmath,amssymb}

\usepackage[normalem]{ulem}

\usepackage{cancel}

\usepackage{hyperref}
\hypersetup{colorlinks=true, linkcolor=red, filecolor=magenta, urlcolor=blue}

\begin{document}

\title{Magnetic Fields from Compensated Isocurvature Perturbations}

\author{Jordan Flitter}
\email{E-mail: jordanf@post.bgu.ac.il}
\affiliation{Physics Department, Ben-Gurion University of the Negev,\\ Beer-Sheva 84105, Israel}
\author{Cyril Creque-Sarbinowski}
\affiliation{Center for Computational Astrophysics, Flatiron Institute,\\
162 Fifth Avenue,
New York, New York 10010}
\author{Marc Kamionkowski}
\affiliation{William H.\ Miller III Department of Physics and Astronomy, Johns Hopkins University, Baltimore, Maryland 21218, USA}
\author{Liang Dai}
\affiliation{Department of Physics, University of California, 366 Physics North MC 7300, Berkeley, CA 94720, USA}

\begin{abstract}
Compensated isocurvature perturbations (CIPs) are perturbations
to the primordial baryon density that are accompanied by
dark-matter-density perturbations so
that the total matter density is unperturbed.  Such CIPs, which may
arise in some multi-field inflationary models, can be
long-lived and only weakly constrained by current
cosmological measurements.  Here we show that the CIP-induced
modulation of the electron number density interacts with the
electron-temperature fluctuation associated with primordial
adiabatic perturbations to produce, via the Biermann-battery
mechanism, a magnetic field in the post-recombinaton Universe.
Assuming the CIP amplitude saturates the current BBN bounds, this magnetic field can be stronger than $10^{-15}\,\mathrm{nG}$ at $z\simeq20$ and stronger by an order of magnitude than
that (produced at second order in the adiabatic-perturbation
amplitude) in the standard cosmological model, and thus can serve as a possible seed for galactic dynamos.
\end{abstract}

\maketitle

\section{Introduction}
Our current cosmological model is consistent with the idea of a
period of inflationary expansion in the early Universe that
generated the primordial density perturbations that seeded the
growth of cosmic large-scale structure.  The canonical model is,
however, no more than a toy model, and so efforts are aimed to
seek new relics in primordial perturbations, beyond the nearly
scale-invariant Gaussian adiabatic perturbations predicted in
the simplest models.  Possibilities include various types of
non-Gaussianity, departures from scale invariance, and assorted
type of isocurvature perturbations
\cite{Baumann:2011nk,Assassi:2012zq,Chen:2012ge,Noumi:2012vr,Arkani-Hamed:2015bza,Lee:2016vti,Kumar:2017ecc,An:2017hlx,An:2017rwo,Baumann:2017jvh,Kumar:2018jxz,Anninos:2019nib,Gong:2013sma,Pi:2012gf}.

Included among these possibilities are compensated isocurvature perturbations
(CIPs), perturbations to the dark-matter density that are
compensated by baryon-density perturbations in such a way that the isocurvature part of
the total matter perturbation vanishes \cite{Gordon:2002gv,Gordon:2009wx,Holder:2009gd}.  CIPs, which can arise
in some multi-field models
\cite{Linde:1996gt,Sasaki:2006kq,Gordon:2009wx,Lyth:2002my,Gordon:2002gv,Langlois:2000ar,He:2015msa}
or during baryogenesis \cite{DeSimone:2016ofp} are particularly intriguing as CIPs
induce (at linear order) no temperature fluctuations in the
cosmic microwave background (CMB).   The fluctuations remain
frozen until shortly after recombination due to CMB drag.  The subsequent evolution is triggered by the
baryon-gas pressure, which is small, thus affecting
perturbations only on very small scales.  Constraints to CIPs on
large scales come from higher-order effects on the CMB power
spectrum
\cite{Munoz:2015fdv,Heinrich:2016gqe,Smith:2017ndr,Valiviita:2017fbx,Planck:2018jri}
and the CMB trispectrum
\cite{Grin:2011tf,Grin:2011nk,Grin:2013uya} while CIPs on small
distance scales may be manifest in CMB spectral
distortions \cite{Haga:2018pdl,Chluba:2013dna} or the
recombination history \cite{Lee:2021bmn}.  Still, these effects
arise only at higher order in perturbation theory and so are fairly
weakly constrained.  The effects of CIPs have also been
considered for baryon acoustic
oscillations~\cite{Soumagnac:2016bjk,Soumagnac:2018atx,Heinrich:2019sxl};
21-cm fluctuations~\cite{Gordon:2009wx}; velocity acoustic
oscillations \cite{Munoz:2019rhi,Munoz:2019fkt} in the 21-cm
power spectrum~\cite{Hotinli:2021xln}; scale-dependent
bias~\cite{Barreira:2019qdl,Barreira:2020lva}; and kSZ
tomography
\cite{Hotinli:2019wdp,Sato-Polito:2020cil,Kumar:2022bly}. 

Here we show that CIPs induce magnetic fields by interacting with primordial adiabatic perturbations
during the dark ages, after recombination but before the epoch
of reionization.  The CIP gives rise to an isothermal
perturbation to the electron number density that then interacts,
via the Biermann-battery mechanism \cite{Naoz:2013wla}, with the
electron-temperature gradients associated with the adiabatic
perturbation to generate a magnetic field. A similar mechanism
operates in the standard cosmological model at second order in the
adiabatic-density-perturbation amplitude \cite{Naoz:2013wla} and
generates magnetic fields weaker than $\mathcal O(10^{-15}\,{\rm nG})$ at the
redshifts, $z\simeq20$, at which the first structures become
nonlinear. As the CIP amplitude may be four orders of magnitude larger than the
adiabatic-perturbation amplitude, one might expect that the CIPs can induce magnetic fields of order $10^{-11}$ nG at redshifts $z\sim20$ and thus possibly detectable by 21-cm measurements \cite{Venumadhav:2014tqa, Gluscevic:2016gns}. Yet, our analysis shows that cancelations in the amplitude of the fluctuations in the electron number-density suppress that enhancement.

The remaining parts of this paper are organized as follows. In Section~\ref{sec: Compensated isocurvature perturbations} we discuss large- and small-scale constraints on the CIP amplitude and relate those to the amplitude of the CIPs-induced free-electron-density fluctuations. We then derive, in Section~\ref{Sec: Biermann battery mechanism}, the magnetic fields that could have arised from the CIPs. We conclude with a discussion on our results in Section~\ref{sec: Discussion}.  

In this work we have adopted the cosmological parameters from the best-fit of Planck 2018~\cite{Planck:2018vyg}, that is a Hubble constant $h=0.6736$, a primordial curvature amplitude $A_s=2.1\times10^{-9}$ with a spectral index $n_s=0.9649$, and total matter and baryons density parameters $\Omega_m=0.3153$, $\Omega_b=0.0493$.

\section{Compensated isocurvature perturbations}\label{sec: Compensated isocurvature perturbations}
We describe the CIP field $\Delta({\bf x}) =\delta \rho_b({\bf
x})/\bar \rho_b$ to be the isocurvature fractional perturbation to the baryon
density.  This field is then accompanied by a fractional isocurvature perturbation
$\delta \rho_c({\bf x})/\bar \rho_c = -f_b \Delta({\bf x})$ to
the cold-dark-matter density, where $f_b=\Omega_b/\Omega_c$ is
the ratio of the baryon and dark-matter densities.  We take the
primordial CIP field to be a realization of a random field with
a power spectrum $\propto k^\alpha$ with the wavenumber $k$.
The time evolution of the CIP field is simple.  The absence of
any density perturbation implies no curvature perturbations (at least at linear order) and
thus no gravitational acceleration.  The pressure gradients in
the baryon-photon fluid introduced by the baryon fluctuation are
extremely small.  The baryon isocurvature perturbation then remains frozen
through radiation drag, which ends at redshift $z\simeq 800$,
when the baryon temperature is roughly 0.2 eV and the baryon
sound speed thus $c_s\sim 1.3\times 10^{-5}\,c$. At this point, the
baryons then spread out at the sound speed, thus smoothing
fluctuations on comoving distance scales smaller than $\sim
2\times 10^{-2}$~Mpc, comparable to the Jeans scale. 

If the power-law index $\alpha>-3$, then the perturbations are
largest at small wavelengths (high $k$).  If the perturbation
amplitude is large, it will affect the agreement between
observed light-element abundances and the predictions of
big-bang nucleosynthesis (BBN).  If the dependence of light-element abundances on the baryon density is perfectly linear, then there will be no change to the abundances after averaging over small-scale fluctuations. The dependence of the deuterium abundance on the baryon density is, however, not linear; it is approximated by $(D/H) \propto \Omega_b^{-3/2}$~\cite{Cyburt:2015mya}.  Tthe deuterium abundance will thus be shifted at quadratic order in $\Delta$ by $\left(1+\Delta\right)^{-3/2}\approx1-\frac{3}{2}\Delta+\frac{15}{8}\Delta^2$,
and after averaging over many small-scale fluctuations the fractional variation in the deuterium abundance is $\delta(D/H)/(\overline{D/H})\sim 2 \langle \Delta^2 \rangle$. The $\sim1\%$ precision of
the current deuterium abundance~\cite{Cooke:2017cwo} then suggests $\langle \Delta^2
\rangle \lesssim 5\times 10^{-3}$.

If on the other hand $\alpha\leq-3$, the baryon density is smooth on small scales but varies on large scales. The rapid Compton interactions prior to recombination will imprint these large scales fluctuations in the baryon density field into large scales fluctuations in the CMB, which would be observed as a difference between the CMB power spectrum on one half of the sky and that on the other half. Current CMB constraints indicate that $\langle \Delta^2
\rangle \lesssim  4\times10^{-3}$~\cite{Planck:2018jri}.

\begin{figure}
\includegraphics[width=\columnwidth]{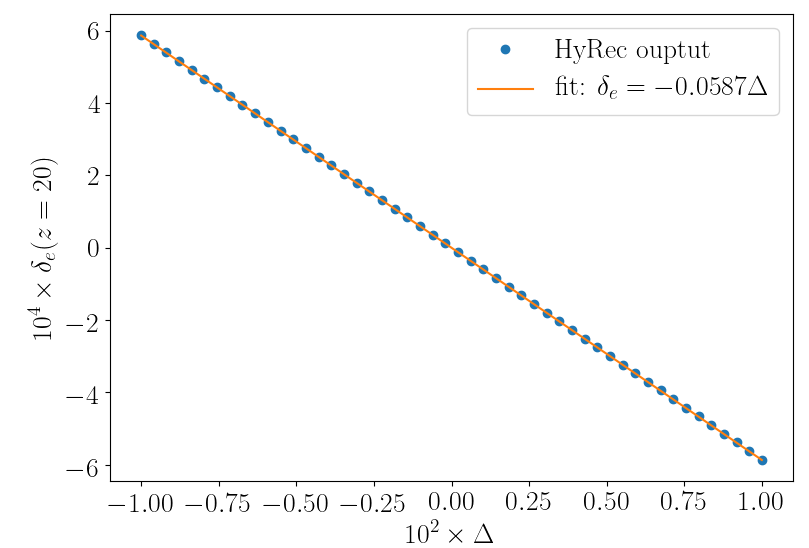}
\caption{Fluctuations in the fractional free-electron-density at $z=20$ as a function of the CIP amplitude. This figure was made by running {\tt{CLASS}}~\cite{Lesgourgues:2011re} (which inherently runs {\tt{HyRec}}) with different values of $\Omega_b$ and $\Omega_c$ that are controlled by the CIP amplitude $\Delta$. As {\tt{CLASS}} and {\tt{HyRec}} return $n_e/n_\mathrm{H}$, where $n_\mathrm{H}$ is the hydrogen-number-density, that quantity was multiplied by $n_\mathrm{H}\propto\Omega_b\left(1-Y_\mathrm{He}\right)$, where $Y_\mathrm{He}$ is the helium-mass-ratio (a quantity that is interpolated by {\tt{CLASS}} given the cosmological parameters).}
\label{fig:delta_e}
\end{figure}

The
Biermann-battery mechanism will depend on the fractional
free-electron-density perturbation $\delta_e(\mathbf{x}) \equiv 
\delta n_e(\mathbf{x})/\bar n_e$.
The free-electron-density $n_e$ at any given point in the post-recombination Universe
will be the product of the baryon density and the free-electron
fraction $x_e$. After recombination, 
the free-electron fraction is proportional
to $\left(1+\Delta\right)^{-1.05}$ (for small values of $\Delta$), an approximation that we have
verified with detailed calculations from {\tt{HyRec}}
\cite{Ali-Haimoud:2010hou,Lee:2020obi}). Due to an $\mathcal O\left(\%1\right)$ correction from the helium abundance~\cite{Cyburt:2015mya}, the induced electron-density
perturbation by the CIPs is therefore
\begin{equation}\label{Eq: delta_e(x)}
\delta_e({\bf x}) \approx -0.06 \Delta({\bf x})\equiv-\xi\Delta({\bf x}), 
\end{equation}
as we have verified numerically with {\tt{HyRec}}, see Fig.~\ref{fig:delta_e}. Ultimately, the suppression by $\xi$ will generate weaker magnetic fields, even though the CIP amplitude can be four orders of magnitude stronger than the adiabatic perturbations.

For simplicity, we will take
the free-electron power spectrum to be in the form of a power-law,
\begin{equation}\label{Eq: Pe}
     P_e\left(k\right) = A_e k_{\rm max}^{-3} (k/k_{\rm max})^{\alpha}\,\Theta(k_{\rm max} - k) \Theta(k-k_{\rm min}),
\end{equation} 
parameterized in terms of an amplitude $A_e$ and power-law
index $\alpha$. For a slow-roll inflation, a scale-invariant CIP power-spectrum is usually considered (with $\alpha=-3$), a choice that is also consistent with the latest Planck CMB analysis~\cite{Planck:2018jri}. Here we work with a general index $\alpha$ to allow comparison of our results with other CIP probes in the literature. In our model, the CIPs power vanishes at Fourier modes below $k_\mathrm{min}$, which essentially corresponds to the horizon scale, and above $k_\mathrm{max}$. For $k_\mathrm{max}$ we shall adopt as our fiducial value the Jeans scale at the end of the drag epoch, i.e. $k_\mathrm{max}\sim400\,\mathrm{Mpc}^{-1}$.

The BBN bound on $\langle \Delta^2 \rangle$ implies an
electron-density variance $\langle \delta_e^2 \rangle =
(2\pi^2)^{-1}\int \, k^2\ dk\,P_e(k) \lesssim 5\times
10^{-3}\,\xi^2$. We thus constrain 
\begin{equation}\label{Eq: BBN bound}
A_e \lesssim (5\times
10^{-3}\,\xi^2)\times
2\pi^2 (3+\alpha)\qquad(\alpha>-3).
\end{equation}
To derive the CMB bounds on $A_e$, we consider the CIP variance, smoothed on
a sphere of radius $R$,
\begin{equation}
\langle\Delta^2\rangle_R=\frac{1}{2\pi^2}\int \,k^2\,dk\left[\frac{3j_1\left(kR\right)}{kR}\right]^2P_\Delta\left(k\right),
\end{equation}
where $j_1\left(x\right)$ is the spherical Bessel function of order $1$. The CIPs power spectrum is obtained from Eq.~\eqref{Eq: delta_e(x)}, $P_\Delta\left(k\right)=\xi^{-2} P_e\left(k\right)$. By taking $R=R_\mathrm{CMB}\sim 125\,\mathrm{Mpc}$~\cite{Smith:2017ndr, Kumar:2022bly}, the CMB constraints on the CIP variance imply
\begin{equation}\label{Eq: CMB bound}
A_e \lesssim\left(4\times
10^{-3}\,\xi^2\right)/ I_\alpha\qquad(\alpha\leq-3),
\end{equation}
where
\begin{equation}
I_\alpha=\frac{1}{2\pi^2}\int_{k_\mathrm{min}}^{k_\mathrm{max}} \,\frac{k^{2+\alpha}\,dk}{k_\mathrm{max}^{3+\alpha}}\left[\frac{3j_1\left(kR_\mathrm{CMB}\right)}{kR_\mathrm{CMB}}\right]^2.
\end{equation}

\section{Biermann battery mechanism}\label{Sec: Biermann battery mechanism}
We now consider the magnetic fields produced in the
post-recombination Universe by the interaction of electron-density
fluctuations, with $k \leq k_{\rm max}$,
with primordial adiabatic density perturbations.
In the standard cosmological model these density perturbations are characterized by the $\Lambda$CDM power spectrum.  Right after recombination, the growth of perturbations to the baryon density are suppressed by Compton drag, but at later times, $z\lesssim 800$, the baryons freely fall and later acquire the same distribution as dark matter, but only for Fourier modes with wavelengths longer than the baryon Jeans scale \cite{Naoz:2010hg}. 
The gas is adiabatic and so baryon-temperature perturbations (and thus electron-temperature perturbations) have an amplitude 2/3 times the density-perturbation amplitude.  This linear-theory evolution proceeds until a redshift $z\sim20$ at which point fluctuations are suppressed at scales smaller than the Jeans scale\footnote{Strictly speaking, the comoving Jeans scale at $z=20$ is $\sim900\,\mathrm{Mpc}^{-1}$. To reduce clutter though, we refer to $k_J \simeq 200$~Mpc$^{-1}$ as the scale where the temperature fluctuations are suppressed by $k^{-2}$ compared to the baryons-density fluctuations~\cite{Naoz:2005pd}.}\footnote{In addition, in our analysis we neglect effects from the relative velocity between baryons and cold-dark-matter $v_\mathrm{bc}$. We anticipate, based on the treatment in Ref.~\cite{Naoz:2013wla}, the including of $v_\mathrm{bc}$ would result an $\mathcal O\left(1\right)$ correction of the induced magnetic field, as well as to an extension of the magntic field power spectrum to larger wavenumbers.} $k_J \simeq 200$~Mpc$^{-1}$.  We thus here calculate the generation of magnetic fields at redshifts $20\lesssim z \lesssim 800$ after baryon drag and before nonlinear structures form.  Nonlinear evolution is likely to amplify the magnetic fields, perhaps considerably, and so the magnetic-field strengths we obtain should be considered conservative lower bounds.

Magnetic fields are generated in the Biermann-battery mechanism if there is a component of the gradient of the electron temperature that is perpendicular to the gradient of the electron density. The evolution of the cosmic magnetic field $\mathbf B$ is related to the electric field $\mathbf E$ through Faraday's law, $\partial\mathbf B/\partial t=-c\mathbf \nabla\times\mathbf E$, with $c$ being the speed of light. Taking the pressure term to be the dominant term in the generalized Ohm's law~\cite{2004ppa..book.....K}, the electric field is $\mathbf E=-\mathbf \nabla p_e/\left(n_ee\right)$, where $p_e$ is the electron pressure and $e$ is the electron charge. Accounting for the expansion of the Universe, the evolution of the magnetic field is then~\cite{Papanikolaou:2023nkx}
\begin{eqnarray}
\nonumber\frac{d}{d t}\left(a^2\mathbf{B}\right)&=&-\frac{c}{en_{e}^{2}}\mathbf{\nabla}n_{e}\times\mathbf{\nabla}p_{e}
\\&=&-\frac{ck_{B}}{en_{e}}\mathbf{\nabla}n_{e}\times\mathbf{\nabla}T,
\end{eqnarray}
where $a\left(t\right)$ is the scale factor and the second line follows the equation of state of collisionless electrons, $p_e=n_ek_BT$, with $k_B$ the Boltzmann constant and $T$ the electron temperature. After defining $\delta_T\equiv\delta T/\bar T$ to be the fractional electron-temperature perturbation and $\bar T$ the mean electron temperature (which we take to be the mean baryon temperature), in the lowest order of perturbation theory we arrive at~\cite{Naoz:2013wla}

\begin{equation}
     \frac{d}{dt} \left(a^2 \mathbf B \right) = - \frac{ c k_B \bar
     T}{e} \mathbf \nabla \delta_e \times \mathbf \nabla \delta_T.
\label{eqn:biermann}     
\end{equation}

The Fourier components $\tilde{\mathbf B}(\mathbf k,t)$ of the magnetic field
induced between some initial time $t_i$ and time $t$ are given
by
\begin{flalign}
&\nonumber\tilde{\mathbf B}(\mathbf k,t)=\frac{c k_B}{a^2\left(t\right)e} \int_{t_i}^{t}\, dt'\, \bar T(t')&
\\&\hspace{5mm}\times\int\frac{d^3 k_1}{(2\pi)^3} \left[ \mathbf k_1 \times (\mathbf k-\mathbf k_1) \right] \tilde\delta_e(\mathbf k_1,t') \tilde\delta_T(\mathbf k - \mathbf k_1,t').&
\end{flalign}
As there is no gravitational attraction in the linear order of the CIP theory, we approximate the electron isocurvature-density fluctuation as constant in time over the relevant wavelengths after recombination, and the electron-temperature perturbation scales as the linear-theory growth factor $D(z)$ (normalized to unity today) which varies as $D(z) \propto 1/(1+z)$ over the relevant redshifts.  The electron temperature $\bar T(z) \propto (1+z)^2$ and the time is $t\simeq (2/3)(\Omega_m H_0^2)^{-1/2}(1+z)^{-3/2}$ at these redshifts.  The redshift (or time) dependence then factorizes and allows us to write the magnetic-field power spectrum at redshift $z$ as
\begin{eqnarray}\label{Eq: P_B(k,z)}
    P_B(k,z) &=& \left[F_B(z) \right]^2 \int \frac{d^3 k_1}{(2\pi)^3} \left[ \mathbf{k}_1 \times (\mathbf{k} - \mathbf{k}_1) \right]^2 \nonumber \\
    & & \times P_e(k_1) P_T\left( |\mathbf{k} - \mathbf{k}_1| \right),
\end{eqnarray}
where
\begin{equation}
     F_B(z) = \frac{2 c k_B \bar T(z)D(z)}{ e \sqrt{\Omega_m} H_0}  (1+z)^{1/2}.
\end{equation}     
The scalings of $\bar T(z)$ and $D(z)$ with $z$ imply that $F_B(z)\propto (1+z)^{3/2}$.  This scaling is slower than the $(1+z)^2$ scaling for a static comoving magnetic field, indicating that the comoving magnetic field is generated primarily at late times.  Our rough treatment of the evolution of the baryon temperature at early times is thus justified and we hereafter adopt $k_\mathrm{min}=3\times10^{-4}\,\mathrm{Mpc}^{-1}$, corresponding to the horizon scale at $z=20$. Numerically, the baryon temperature is $\bar T(z=20) \simeq 10$ K, and $D(z=20) \simeq 0.06$, and then
\begin{equation}
     F_B(z) \simeq 4.1\times10^{-27}\, {\rm G}\,{\rm Mpc}^2\, \left(\frac{1+z}{21}\right)^{3/2}.
\end{equation}

The magnetic-field variance then becomes
\begin{flalign}
  &\left\langle \mathbf{B}^2 \right\rangle = \int \frac{d^3k}{(2\pi)^3} P_B(k)& \nonumber \\
        &\hspace{8mm}=  [F_B(z)]^2 \langle \sin^2 \theta \rangle& \nonumber \\
        &\hspace{12mm}\times \left[ \int \frac{d^3k}{(2\pi)^3} k^2 P_e(k) \right] \left[ \int \frac{d^3k}{(2\pi)^3} k^2 P_T(k) \right],&
\label{eqn:rhoB}        
\end{flalign}
where $\langle\sin^2\theta \rangle=2/3$ is the angle between $\mathbf{k}_1$ and $\mathbf{k}$ averaged over all directions.  The first integral in Eq.~(\ref{eqn:rhoB}) evaluates to $A_e k_{\rm max}^2 [2\pi^2 (5+\alpha)]^{-1}\left| 1- (k_{\rm min}/k_{\rm max})^{\alpha+5} \right|$.  We evaluate the second integral using {\tt CLASS} \cite{Lesgourgues:2011re}, cutting off the integral at the Jeans scale $k_J \simeq200$~Mpc.  It comes out to $2.2\times 10^5\, (k_J/200\, {\rm Mpc}^{-1})^2$~Mpc$^{-2}$, where the scaling with $k_J$ arises given that the integral is dominated by the high-$k$ limit where $P_m(k)$ varies very slowly with $k$. 

We thus find an rms magnetic-field strength 
\begin{eqnarray}\label{Eq: Brms}
      \left\langle \mathbf{B}^2 \right\rangle^{1/2} &\simeq &2.9 \times 10^{-15} \, {\rm nG}\, \left(\frac{A_e}{\left(2\times10^{-5}\right)2\pi^2 |5+\alpha|}\right)^{1/2} \nonumber \\ 
      & & \times\frac{k_J}{200\,{\rm Mpc}^{-1}}\frac{k_{\rm max}}{400\,{\rm Mpc}^{-1}} \left(\frac{1+z}{21}\right)^{3/2} \nonumber \\
      & & \times \left| 1 - \left( \frac{k_{\rm min}}{k_{\rm max}} \right)^{5+\alpha} \right|^{1/2},
\end{eqnarray}
at redshifts $z\simeq 20$.
A few comments:  (1) The scaling with redshift is expected to break down for redshifts $z\lesssim20$ for several reasons.  First, small-scale perturbations will go nonlinear, violating the assumption of linearity.  Second, the formation of stars will heat the gas and increase the temperature.  Moreover, we expect that the motions of  magnetized gas that winds up in gravitationally bound systems will lead, through the dynamo mechanism, to magnetic fields within halos that are far stronger than the seed fields provided by our analysis.  (2) The behavior of the small-scale power spectrum, on scales smaller than the Jeans scale, can be calculated, as we show in the appendix.  Here we have modeled it as a strict cutoff for simplicity and to help indicate the uncertainty on this small-scale physics. Yet, we find that Eq.~\eqref{Eq: Brms} is accurate to the order of $\mathcal O\left(10\%\right)$ when we compare it to numerical calculations with a more refined modeling of the suppression at small scales, as we discuss next.

\begin{figure}
\includegraphics[width=\columnwidth]{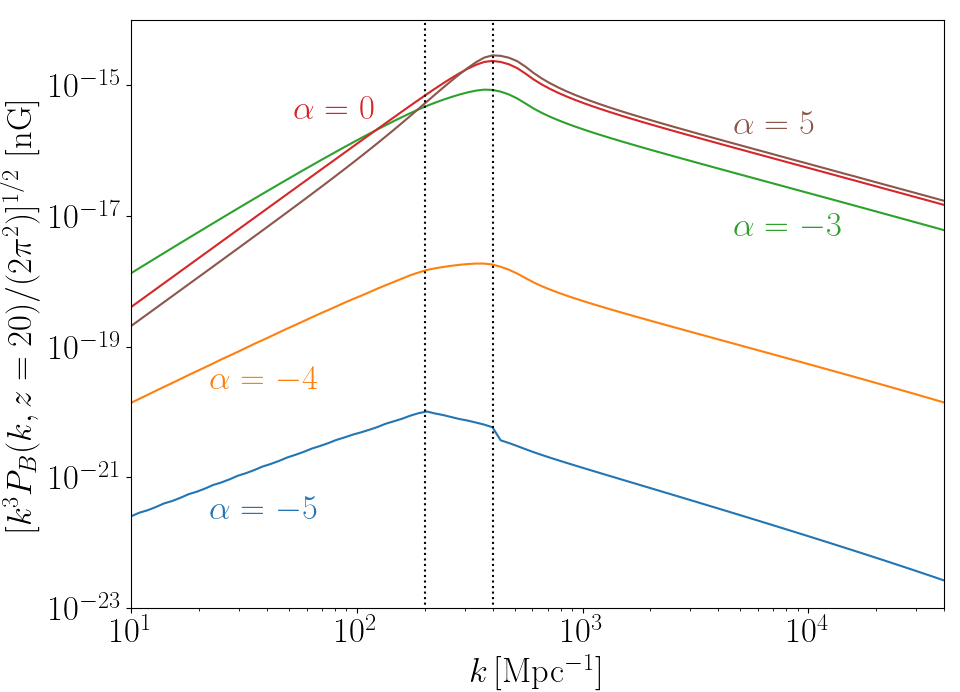}
\caption{The magnetic-field power spectrum as a function of
wavenumber for different values of the electron-density
spectral index $\alpha$. For values of $\alpha>-3$ ($\alpha\leq-3$), the
normalization of the CIP power spectrum is taken to be the
maximum allowed by the BBN (CMB) constraint, Eq.~\eqref{Eq: BBN bound} (Eq.~\eqref{Eq: CMB bound}).  The vertical lines
indicate the assumed Jeans wavenumber $k_J=200$ Mpc$^{-1}$ and
the cutoff $k_{\rm max}=400$ Mpc$^{-1}$ in the electron-density
power spectrum.}
\label{fig:Pb}
\end{figure}

\section{Discussion}\label{sec: Discussion}

In Fig.~\ref{fig:Pb} we plot the magnetic-field power spectrum for values of $-5  \leq \alpha \leq 5$, in each case taking $A_e$ to be the maximum value allowed by the BBN constraint for $\alpha>-3$ (Eq.~\ref{Eq: BBN bound}) and by the CMB  bound for $\alpha<-3$ (Eq.~\ref{Eq: CMB bound}).   As above, the CIP power spectrum is assumed here to be cut off at $k_{\rm max}=400$~Mpc$^{-1}$; this cutoff gives rise to the sharp drop in $P_B(k)$ at this value of $k$, mostly evident at $\alpha=-5$.  In this calculation, though, we use the standard $\Lambda$CDM power spectrum from {\tt CLASS} \cite{Lesgourgues:2011re}, with best-fit Planck cosmological parameters \cite{Planck:2018vyg}.  We then extend it to smaller scales (higher $k$) with the BBKS approximation \cite{Bardeen:1985tr} with the baryon correction of Ref.~\cite{Sugiyama:1994ed}.  We then continue the matter power spectrum to $k>k_J$, above the Jeans scale, with a suppression $(k/k_J)^{-4}$ \cite{Naoz:2005pd, Naoz:2010hg}, for $k>k_J=200$~Mpc$^{-1}$, relative to the BBKS power spectrum.  

The magnetic-field power peaks in all cases at $k\simeq k_{\rm
max}$, but is a bit more broadly distributed to lower $k$ for
$\alpha<-3$ (see derivation in the appendix for the power-law behavior at small and large scales). For $\alpha>0$ the power at the vicinity of $k\sim k_\mathrm{max}$ surpasses $10^{-15}\,\mathrm{nG}$. In all cases, the magnetic-field energy density is
negligible compared with the thermal energy density in the gas
at these redshifts (this would require $B\sim$~nG), and so it is
dynamically unimportant.  It also follows, from this figure,
that this mechanism is not constrained by upper limits of
$\sim$~nG to intergalactic magnetic fields
\cite{Amaral:2021mly,Pandey:2012ss,Chongchitnan:2013vpa} nor
strong enough to be relevant for the magnetic fields suggested
by interpretation of variability of gamma-ray sources
\cite{AlvesBatista:2020oio,Fermi-LAT:2018jdy,MAGIC:2022piy},
which reach as small as $B\sim10^{-8}$~nG. If the CIP amplitude is close to its BBN upper bounds, as we have assumed in this calculation, the field strengths
may be suitable to account for seed fields for galactic dynamos
\cite{Widrow:2002ud}, which in some models can be as small as $10^{-21}\,\mathrm{nG}$ at the time of galaxy formation~\cite{Davis:1999bt}. The magnetic-field strengths we obtain
can be higher than those that arise at
quadratic order in the primordial density perturbation
\cite{Naoz:2013wla}.

In conclusion, in this work we have studied the spectrum and strength of magnetic fields that could have been generated via the Biermann-battery mechanism, where the fluctuations in the electron number-density are sourced by the CIP field. Our main result is Eq.~\eqref{Eq: Brms}. As a first study on this effect, we favored simplicity over precision in order to estimate the strength of the CIP-induced magnetic fields. Although the analysis presented in this paper can be improved by performing a full perturbation analysis that includes also the relative velocity between baryons and cold-dark-matter, as was done in Ref.~\cite{Naoz:2013wla}, we do not expect these improvements to alter the qualitative features of the magnetic fields we obtained, certainly not by orders of magnitude.

At first, one may have surmised that CIP-induced magnetic fields could be up to a few orders of magnitude larger, given that the CIP amplitude can be orders of magnitude larger than the adiabatic-perturbation amplitude.  Much of that gain is erased, however, by the near cancellation between the fluctuations in the free-electron fraction and the baryon density in the CIP.  Our calculation shows that, when all the dust has settled, the CIP allows for a larger magnetic field, but not much.

\begin{acknowledgments}
We thank Dan Grin and Yacine Ali-Ha\"{i}moud for useful discussions. We would also like to thank the anonymous referee for useful comments that improved the quality of the paper. JF is supported by
the Zin fellowship awarded by the BGU Kreitmann School. CCS was
supported at Johns Hopkins by the NSF Graduate Research
Fellowship under Grant No.\ DGE1746891 and the Bill and Melinda
Gates Foundation. This work was supported at Johns Hopkins by
NSF Grant No.\ 2112699 and the Simons Foundation.
LD acknowledges research grant support from the Alfred P. Sloan Foundation (Award Number FG-2021-16495). 
\end{acknowledgments}

\appendix
\section*{Appendix: Analytical approximations for the magnetic field power spectrum}
Eq.~\eqref{Eq: P_B(k,z)} can written in the following form,
\begin{equation}\label{Eq: PB with I(k)}
P_B\left(k\right)=\left[F_B\left(z\right)\right]^2I\left(k\right),
\end{equation}
where
\begin{flalign}\label{Eq: I(k)}
&\nonumber I\left(k\right)=\int \frac{d^3 k_1}{(2\pi)^3}\left|\mathbf k_1 \times \mathbf k \right|^2 P_e\left( k_1 \right)P_T \left(| \mathbf k - \mathbf k_1 | \right)&
\\&\hspace{7mm}=\frac{A_ek^{7+\alpha}}{4\pi^2k_\mathrm{max}^{3+\alpha}}\int_{-1}^1 d\mu\left(1-\mu^2\right)I_x\left(k,\mu\right),&
\end{flalign}
and
\begin{equation}\label{Eq: Ix}
I_x\left(k,\mu\right)=\int_{x_\mathrm{min}}^{x_\mathrm{max}}dx\,x^{4+\alpha}P_T \left[k\beta\left(x,\mu\right)\right],
\end{equation}
where $\beta\left(x,\mu\right)=\sqrt{1+x^2-2x\mu}$, $x_\mathrm{min}\equiv k_\mathrm{min}/k$ and $x_\mathrm{max}\equiv k_\mathrm{max}/k$. For simplicity, we model the temperature power spectrum as follows
\begin{equation}
P_T\left(k\right)=\frac{4}{9}P_m\left(k\right)\times\begin{cases}
1 & k\leq k_J \\
x_J^{4} & k\geq k_J
\end{cases},
\end{equation}
where $P_m\left(k\right)$ is the linear matter power spectrum and $x_J\equiv k_J/k$.

Below we examine $I\left(k\right)$ in two limits. Throughout the derivation, we assume $k\gg k_\mathrm{eq}\approx0.01\,\mathrm{Mpc}^{-1}$ and assume that the matter power spectrum scales as $P_m\left(k\right)\propto k^{n_s-4}$ (we ignore small logarithmic corrections). We also assume that $k_J$ and $k_\mathrm{max}$ are of the same order and limit ourselves to $\alpha\geq -5$. The final expressions for $I\left(k\right)$, Eqs.~\eqref{Eq: I(k) low} and \eqref{Eq: I(k) high} capture very well the power-laws behavior seen in Fig.~\ref{fig:Pb} and provide a good order of magnitude estimation.

\subsection{First limit: \texorpdfstring{$k\ll k_J<k_\mathrm{max}$ ($1\ll x_J < x_\mathrm{max}$)}{}}
In the limit where $x_\mathrm{max}\gg 1$, we split the integral of $I_x\left(k,\mu\right)$ into two regimes, $I_x\left(k,\mu\right)=I_x^{x_\mathrm{min}\to1}\left(k,\mu\right)+I_x^{1\to x_\mathrm{max}}\left(k,\mu\right)$. For the first piece, $I_x^{x_\mathrm{min}\to1}\left(k,\mu\right)$, we approximate $\beta\left(x,\mu\right)\approx1$, which yields
\begin{eqnarray}
\nonumber I_x^{x_\mathrm{min}\to1}\left(k,\mu\right)&=&\frac{4}{9}P_m\left(k\right)C_\alpha^{(1)}\left(k\right)
\\&\approx&\frac{4}{9}P_m\left(k_\mathrm{max}\right)C_\alpha^{(1)}\left(k\right)x_\mathrm{max}^{4-n_{s}}.
\end{eqnarray}
where $C_\alpha^{(1)}\left(k\right)$ is an $\mathcal O\left(1\right)$ factor (assuming $\alpha\geq-5$)
\begin{equation}
C_\alpha^{(1)}\left(k\right)=\begin{cases}
\frac{1-x_\mathrm{min}^{5+\alpha}}{5+\alpha}\approx\frac{1}{5+\alpha} & \alpha+5\neq0 \\
-\ln x_\mathrm{min} & \alpha+5=0
\end{cases}.
\end{equation}
For the second piece, $I_x^{1\to x_\mathrm{max}}\left(k,\mu\right)$, we approximate $\beta\left(x,\mu\right)\approx x$.
Thus, in the range $1\leq x\leq x_\mathrm{max}$,
\begin{flalign}
\nonumber &P_T \left[k\beta\left(x,\mu\right)\right]\approx\frac{4}{9}P_m\left(k_\mathrm{max}\right)\left(\frac{x}{x_\mathrm{max}}\right)^{n_s-4}&
\\&\hspace{30mm}\times\begin{cases}
1 & 1\leq x\leq x_J \\
\left(x/x_J\right)^{-4} & x_J\leq x\leq x_\mathrm{max}
\end{cases},&
\end{flalign}
Then, the calculation of $I_x^{1\to x_\mathrm{max}}\left(k,\mu\right)$ is straight-forward,
\begin{flalign}
\nonumber&I_x^{1\to x_\mathrm{max}}\left(k,\mu\right)=\frac{4}{9}P_m\left(k_\mathrm{max}\right)C^{(2)}_\alpha&
\\&\hspace{30mm}\times\begin{cases}
x_\mathrm{max}^{\alpha+5} & \alpha+n_s+1\geq 0 \\
x_\mathrm{max}^{4-n_s} & \alpha+n_s+1< 0
\end{cases},&
\end{flalign}
where $C^{(2)}_\alpha$ is another $\mathcal O\left(1\right)$ constant, 
\begin{equation}
C^{(2)}_\alpha=\begin{cases}
\frac{\left(\alpha+n_{s}+1\right)x_{\mathrm{max}/J}^{\alpha+n_{s}-3}-4}{\left(\alpha+n_{s}+1\right)\left(\alpha+n_{s}-3\right)}x_{\mathrm{max}/J}^{\alpha+n_{s}+1} & \alpha+n_s+1 > 0 \\
\ln x_{J}+\frac{1-x_{\mathrm{max}/J}^{-4}}{4} & \alpha+n_s+1 = 0 \\
-\frac{1}{\alpha+n_{s}+1} & \alpha+n_s+1 < 0
\end{cases},
\end{equation}
where $x_{\mathrm{max}/J}\equiv x_\mathrm{max}/x_J=\mathcal O\left(1\right)$.

Because $x_\mathrm{max}\gg 1$, $x_\mathrm{max}^{\alpha+5}\gg x_\mathrm{max}^{4-n_s}$ for $\alpha+n_s+1\geq 0$ and therefore $I_x^{1\to x_\mathrm{max}}\gg I_x^{x_\mathrm{min}\to1}$, while $I_x^{1\to x_\mathrm{max}}$ and $I_x^{x_\mathrm{min}\to1}$ are comparable for $\alpha+n_s+1<0$. Thus 
\begin{flalign}
\nonumber&I_x\left(k,\mu\right)=\frac{4}{9}P_m\left(k_\mathrm{max}\right)&
\\&\hspace{15mm}\times\begin{cases}
C^{(2)}_\alpha x_\mathrm{max}^{\alpha+5} & \alpha+n_s+1\geq 0 \\
\left(C^{(1)}_\alpha\left(k\right)+C^{(2)}_\alpha\right)x_\mathrm{max}^{4-n_s} & \alpha+n_s+1< 0
\end{cases},&
\end{flalign}
Since $I_x\left(k,\mu\right)$ does not depend on $\mu$, the $\mu$ integral in Eq.~\eqref{Eq: I(k)} gives $4/3$, and we have
\begin{flalign}\label{Eq: I(k) low}
\nonumber&I\left(k\right)=A_B&
\\&\hspace{5mm}\times\begin{cases}
C^{(2)}_\alpha\left(\frac{k}{k_\mathrm{max}}\right)^{2} & \alpha+n_s+1\geq 0 \\
\left(C^{(1)}_\alpha\left(k\right)+C^{(2)}_\alpha\right)\left(\frac{k}{k_\mathrm{max}}\right)^{\alpha+n_s+3} & \alpha+n_s+1< 0
\end{cases},&
\end{flalign}
where
\begin{flalign}
\nonumber &A_B\equiv\frac{4A_eP_m\left(k_\mathrm{max}\right)k_\mathrm{max}^4}{27\pi^2}&
\\&\hspace{5mm}\approx2\times10^{-6}\left(\frac{A_{e}}{10^{-4}}\right)\left(\frac{k_{\mathrm{max}}}{400\,\mathrm{Mpc}^{-1}}\right)^{n_{s}}\,\mathrm{Mpc}^{-1}.&
\end{flalign}

\subsection{Second limit: \texorpdfstring{$k_J< k_\mathrm{max}\ll k$ ($x_J<x_\mathrm{max}\ll 1$)}{}}

In the limit where $x_\mathrm{max}\ll 1$, we approximate $\beta\left(x,\mu\right)\approx1$ and therefore
\begin{flalign}
\nonumber &I_x\left(k,\mu\right)=\frac{4}{9}P_m\left(k\right)x_J^{4}C_\alpha^{(1)}\left(k_\mathrm{max}\right)x_\mathrm{max}^{5+\alpha}&
\\&\hspace{12mm}\approx\frac{4}{9}P_m\left(k_\mathrm{max}\right)x_{\mathrm{max}/J}^{-4}C_\alpha^{(1)}\left(k_\mathrm{max}\right)x_\mathrm{max}^{13+\alpha-n_{s}}.&
\end{flalign}
Again, the $\mu$ integral in Eq.~\eqref{Eq: I(k)} gives $4/3$, and we have
\begin{equation}\label{Eq: I(k) high} I\left(k\right)=A_BC_\alpha^{(1)}\left(k_\mathrm{max}\right)x_{\mathrm{max}/J}^{-4}\left(\frac{k}{k_\mathrm{max}}\right)^{n_s-6}.
\end{equation}

\bibliographystyle{apsrev4-1}
\bibliography{B_from_cips.bib}

\begin{thebibliography}{68}%
\makeatletter
\providecommand \@ifxundefined [1]{%
 \@ifx{#1\undefined}
}%
\providecommand \@ifnum [1]{%
 \ifnum #1\expandafter \@firstoftwo
 \else \expandafter \@secondoftwo
 \fi
}%
\providecommand \@ifx [1]{%
 \ifx #1\expandafter \@firstoftwo
 \else \expandafter \@secondoftwo
 \fi
}%
\providecommand \natexlab [1]{#1}%
\providecommand \enquote  [1]{``#1''}%
\providecommand \bibnamefont  [1]{#1}%
\providecommand \bibfnamefont [1]{#1}%
\providecommand \citenamefont [1]{#1}%
\providecommand \href@noop [0]{\@secondoftwo}%
\providecommand \href [0]{\begingroup \@sanitize@url \@href}%
\providecommand \@href[1]{\@@startlink{#1}\@@href}%
\providecommand \@@href[1]{\endgroup#1\@@endlink}%
\providecommand \@sanitize@url [0]{\catcode `\\12\catcode `\$12\catcode
  `\&12\catcode `\#12\catcode `\^12\catcode `\_12\catcode `\%12\relax}%
\providecommand \@@startlink[1]{}%
\providecommand \@@endlink[0]{}%
\providecommand \url  [0]{\begingroup\@sanitize@url \@url }%
\providecommand \@url [1]{\endgroup\@href {#1}{\urlprefix }}%
\providecommand \urlprefix  [0]{URL }%
\providecommand \Eprint [0]{\href }%
\providecommand \doibase [0]{http://dx.doi.org/}%
\providecommand \selectlanguage [0]{\@gobble}%
\providecommand \bibinfo  [0]{\@secondoftwo}%
\providecommand \bibfield  [0]{\@secondoftwo}%
\providecommand \translation [1]{[#1]}%
\providecommand \BibitemOpen [0]{}%
\providecommand \bibitemStop [0]{}%
\providecommand \bibitemNoStop [0]{.\EOS\space}%
\providecommand \EOS [0]{\spacefactor3000\relax}%
\providecommand \BibitemShut  [1]{\csname bibitem#1\endcsname}%
\let\auto@bib@innerbib\@empty
\bibitem [{\citenamefont {Baumann}\ and\ \citenamefont
  {Green}(2012)}]{Baumann:2011nk}%
  \BibitemOpen
  \bibfield  {author} {\bibinfo {author} {\bibfnamefont {D.}~\bibnamefont
  {Baumann}}\ and\ \bibinfo {author} {\bibfnamefont {D.}~\bibnamefont
  {Green}},\ }\href {\doibase 10.1103/PhysRevD.85.103520} {\bibfield  {journal}
  {\bibinfo  {journal} {Phys. Rev. D}\ }\textbf {\bibinfo {volume} {85}},\
  \bibinfo {pages} {103520} (\bibinfo {year} {2012})},\ \Eprint
  {http://arxiv.org/abs/1109.0292} {arXiv:1109.0292 [hep-th]} \BibitemShut
  {NoStop}%
\bibitem [{\citenamefont {Assassi}\ \emph {et~al.}(2012)\citenamefont
  {Assassi}, \citenamefont {Baumann},\ and\ \citenamefont
  {Green}}]{Assassi:2012zq}%
  \BibitemOpen
  \bibfield  {author} {\bibinfo {author} {\bibfnamefont {V.}~\bibnamefont
  {Assassi}}, \bibinfo {author} {\bibfnamefont {D.}~\bibnamefont {Baumann}}, \
  and\ \bibinfo {author} {\bibfnamefont {D.}~\bibnamefont {Green}},\ }\href
  {\doibase 10.1088/1475-7516/2012/11/047} {\bibfield  {journal} {\bibinfo
  {journal} {JCAP}\ }\textbf {\bibinfo {volume} {11}},\ \bibinfo {pages} {047}
  (\bibinfo {year} {2012})},\ \Eprint {http://arxiv.org/abs/1204.4207}
  {arXiv:1204.4207 [hep-th]} \BibitemShut {NoStop}%
\bibitem [{\citenamefont {Chen}\ and\ \citenamefont
  {Wang}(2012)}]{Chen:2012ge}%
  \BibitemOpen
  \bibfield  {author} {\bibinfo {author} {\bibfnamefont {X.}~\bibnamefont
  {Chen}}\ and\ \bibinfo {author} {\bibfnamefont {Y.}~\bibnamefont {Wang}},\
  }\href {\doibase 10.1088/1475-7516/2012/09/021} {\bibfield  {journal}
  {\bibinfo  {journal} {JCAP}\ }\textbf {\bibinfo {volume} {09}},\ \bibinfo
  {pages} {021} (\bibinfo {year} {2012})},\ \Eprint
  {http://arxiv.org/abs/1205.0160} {arXiv:1205.0160 [hep-th]} \BibitemShut
  {NoStop}%
\bibitem [{\citenamefont {Noumi}\ \emph {et~al.}(2013)\citenamefont {Noumi},
  \citenamefont {Yamaguchi},\ and\ \citenamefont {Yokoyama}}]{Noumi:2012vr}%
  \BibitemOpen
  \bibfield  {author} {\bibinfo {author} {\bibfnamefont {T.}~\bibnamefont
  {Noumi}}, \bibinfo {author} {\bibfnamefont {M.}~\bibnamefont {Yamaguchi}}, \
  and\ \bibinfo {author} {\bibfnamefont {D.}~\bibnamefont {Yokoyama}},\ }\href
  {\doibase 10.1007/JHEP06(2013)051} {\bibfield  {journal} {\bibinfo  {journal}
  {JHEP}\ }\textbf {\bibinfo {volume} {06}},\ \bibinfo {pages} {051} (\bibinfo
  {year} {2013})},\ \Eprint {http://arxiv.org/abs/1211.1624} {arXiv:1211.1624
  [hep-th]} \BibitemShut {NoStop}%
\bibitem [{\citenamefont {Arkani-Hamed}\ and\ \citenamefont
  {Maldacena}(2015)}]{Arkani-Hamed:2015bza}%
  \BibitemOpen
  \bibfield  {author} {\bibinfo {author} {\bibfnamefont {N.}~\bibnamefont
  {Arkani-Hamed}}\ and\ \bibinfo {author} {\bibfnamefont {J.}~\bibnamefont
  {Maldacena}},\ }\href@noop {} {\  (\bibinfo {year} {2015})},\ \Eprint
  {http://arxiv.org/abs/1503.08043} {arXiv:1503.08043 [hep-th]} \BibitemShut
  {NoStop}%
\bibitem [{\citenamefont {Lee}\ \emph {et~al.}(2016)\citenamefont {Lee},
  \citenamefont {Baumann},\ and\ \citenamefont {Pimentel}}]{Lee:2016vti}%
  \BibitemOpen
  \bibfield  {author} {\bibinfo {author} {\bibfnamefont {H.}~\bibnamefont
  {Lee}}, \bibinfo {author} {\bibfnamefont {D.}~\bibnamefont {Baumann}}, \ and\
  \bibinfo {author} {\bibfnamefont {G.~L.}\ \bibnamefont {Pimentel}},\ }\href
  {\doibase 10.1007/JHEP12(2016)040} {\bibfield  {journal} {\bibinfo  {journal}
  {JHEP}\ }\textbf {\bibinfo {volume} {12}},\ \bibinfo {pages} {040} (\bibinfo
  {year} {2016})},\ \Eprint {http://arxiv.org/abs/1607.03735} {arXiv:1607.03735
  [hep-th]} \BibitemShut {NoStop}%
\bibitem [{\citenamefont {Kumar}\ and\ \citenamefont
  {Sundrum}(2018)}]{Kumar:2017ecc}%
  \BibitemOpen
  \bibfield  {author} {\bibinfo {author} {\bibfnamefont {S.}~\bibnamefont
  {Kumar}}\ and\ \bibinfo {author} {\bibfnamefont {R.}~\bibnamefont
  {Sundrum}},\ }\href {\doibase 10.1007/JHEP05(2018)011} {\bibfield  {journal}
  {\bibinfo  {journal} {JHEP}\ }\textbf {\bibinfo {volume} {05}},\ \bibinfo
  {pages} {011} (\bibinfo {year} {2018})},\ \Eprint
  {http://arxiv.org/abs/1711.03988} {arXiv:1711.03988 [hep-ph]} \BibitemShut
  {NoStop}%
\bibitem [{\citenamefont {An}\ \emph {et~al.}(2018{\natexlab{a}})\citenamefont
  {An}, \citenamefont {McAneny}, \citenamefont {Ridgway},\ and\ \citenamefont
  {Wise}}]{An:2017hlx}%
  \BibitemOpen
  \bibfield  {author} {\bibinfo {author} {\bibfnamefont {H.}~\bibnamefont
  {An}}, \bibinfo {author} {\bibfnamefont {M.}~\bibnamefont {McAneny}},
  \bibinfo {author} {\bibfnamefont {A.~K.}\ \bibnamefont {Ridgway}}, \ and\
  \bibinfo {author} {\bibfnamefont {M.~B.}\ \bibnamefont {Wise}},\ }\href
  {\doibase 10.1007/JHEP06(2018)105} {\bibfield  {journal} {\bibinfo  {journal}
  {JHEP}\ }\textbf {\bibinfo {volume} {06}},\ \bibinfo {pages} {105} (\bibinfo
  {year} {2018}{\natexlab{a}})},\ \Eprint {http://arxiv.org/abs/1706.09971}
  {arXiv:1706.09971 [hep-ph]} \BibitemShut {NoStop}%
\bibitem [{\citenamefont {An}\ \emph {et~al.}(2018{\natexlab{b}})\citenamefont
  {An}, \citenamefont {McAneny}, \citenamefont {Ridgway},\ and\ \citenamefont
  {Wise}}]{An:2017rwo}%
  \BibitemOpen
  \bibfield  {author} {\bibinfo {author} {\bibfnamefont {H.}~\bibnamefont
  {An}}, \bibinfo {author} {\bibfnamefont {M.}~\bibnamefont {McAneny}},
  \bibinfo {author} {\bibfnamefont {A.~K.}\ \bibnamefont {Ridgway}}, \ and\
  \bibinfo {author} {\bibfnamefont {M.~B.}\ \bibnamefont {Wise}},\ }\href
  {\doibase 10.1103/PhysRevD.97.123528} {\bibfield  {journal} {\bibinfo
  {journal} {Phys. Rev. D}\ }\textbf {\bibinfo {volume} {97}},\ \bibinfo
  {pages} {123528} (\bibinfo {year} {2018}{\natexlab{b}})},\ \Eprint
  {http://arxiv.org/abs/1711.02667} {arXiv:1711.02667 [hep-ph]} \BibitemShut
  {NoStop}%
\bibitem [{\citenamefont {Baumann}\ \emph {et~al.}(2018)\citenamefont
  {Baumann}, \citenamefont {Goon}, \citenamefont {Lee},\ and\ \citenamefont
  {Pimentel}}]{Baumann:2017jvh}%
  \BibitemOpen
  \bibfield  {author} {\bibinfo {author} {\bibfnamefont {D.}~\bibnamefont
  {Baumann}}, \bibinfo {author} {\bibfnamefont {G.}~\bibnamefont {Goon}},
  \bibinfo {author} {\bibfnamefont {H.}~\bibnamefont {Lee}}, \ and\ \bibinfo
  {author} {\bibfnamefont {G.~L.}\ \bibnamefont {Pimentel}},\ }\href {\doibase
  10.1007/JHEP04(2018)140} {\bibfield  {journal} {\bibinfo  {journal} {JHEP}\
  }\textbf {\bibinfo {volume} {04}},\ \bibinfo {pages} {140} (\bibinfo {year}
  {2018})},\ \Eprint {http://arxiv.org/abs/1712.06624} {arXiv:1712.06624
  [hep-th]} \BibitemShut {NoStop}%
\bibitem [{\citenamefont {Kumar}\ and\ \citenamefont
  {Sundrum}(2019)}]{Kumar:2018jxz}%
  \BibitemOpen
  \bibfield  {author} {\bibinfo {author} {\bibfnamefont {S.}~\bibnamefont
  {Kumar}}\ and\ \bibinfo {author} {\bibfnamefont {R.}~\bibnamefont
  {Sundrum}},\ }\href {\doibase 10.1007/JHEP04(2019)120} {\bibfield  {journal}
  {\bibinfo  {journal} {JHEP}\ }\textbf {\bibinfo {volume} {04}},\ \bibinfo
  {pages} {120} (\bibinfo {year} {2019})},\ \Eprint
  {http://arxiv.org/abs/1811.11200} {arXiv:1811.11200 [hep-ph]} \BibitemShut
  {NoStop}%
\bibitem [{\citenamefont {Anninos}\ \emph {et~al.}(2019)\citenamefont
  {Anninos}, \citenamefont {De~Luca}, \citenamefont {Franciolini},
  \citenamefont {Kehagias},\ and\ \citenamefont {Riotto}}]{Anninos:2019nib}%
  \BibitemOpen
  \bibfield  {author} {\bibinfo {author} {\bibfnamefont {D.}~\bibnamefont
  {Anninos}}, \bibinfo {author} {\bibfnamefont {V.}~\bibnamefont {De~Luca}},
  \bibinfo {author} {\bibfnamefont {G.}~\bibnamefont {Franciolini}}, \bibinfo
  {author} {\bibfnamefont {A.}~\bibnamefont {Kehagias}}, \ and\ \bibinfo
  {author} {\bibfnamefont {A.}~\bibnamefont {Riotto}},\ }\href {\doibase
  10.1088/1475-7516/2019/04/045} {\bibfield  {journal} {\bibinfo  {journal}
  {JCAP}\ }\textbf {\bibinfo {volume} {04}},\ \bibinfo {pages} {045} (\bibinfo
  {year} {2019})},\ \Eprint {http://arxiv.org/abs/1902.01251} {arXiv:1902.01251
  [hep-th]} \BibitemShut {NoStop}%
\bibitem [{\citenamefont {Gong}\ \emph {et~al.}(2013)\citenamefont {Gong},
  \citenamefont {Pi},\ and\ \citenamefont {Sasaki}}]{Gong:2013sma}%
  \BibitemOpen
  \bibfield  {author} {\bibinfo {author} {\bibfnamefont {J.-O.}\ \bibnamefont
  {Gong}}, \bibinfo {author} {\bibfnamefont {S.}~\bibnamefont {Pi}}, \ and\
  \bibinfo {author} {\bibfnamefont {M.}~\bibnamefont {Sasaki}},\ }\href
  {\doibase 10.1088/1475-7516/2013/11/043} {\bibfield  {journal} {\bibinfo
  {journal} {JCAP}\ }\textbf {\bibinfo {volume} {11}},\ \bibinfo {pages} {043}
  (\bibinfo {year} {2013})},\ \Eprint {http://arxiv.org/abs/1306.3691}
  {arXiv:1306.3691 [hep-th]} \BibitemShut {NoStop}%
\bibitem [{\citenamefont {Pi}\ and\ \citenamefont {Sasaki}(2012)}]{Pi:2012gf}%
  \BibitemOpen
  \bibfield  {author} {\bibinfo {author} {\bibfnamefont {S.}~\bibnamefont
  {Pi}}\ and\ \bibinfo {author} {\bibfnamefont {M.}~\bibnamefont {Sasaki}},\
  }\href {\doibase 10.1088/1475-7516/2012/10/051} {\bibfield  {journal}
  {\bibinfo  {journal} {JCAP}\ }\textbf {\bibinfo {volume} {10}},\ \bibinfo
  {pages} {051} (\bibinfo {year} {2012})},\ \Eprint
  {http://arxiv.org/abs/1205.0161} {arXiv:1205.0161 [hep-th]} \BibitemShut
  {NoStop}%
\bibitem [{\citenamefont {Gordon}\ and\ \citenamefont
  {Lewis}(2003)}]{Gordon:2002gv}%
  \BibitemOpen
  \bibfield  {author} {\bibinfo {author} {\bibfnamefont {C.}~\bibnamefont
  {Gordon}}\ and\ \bibinfo {author} {\bibfnamefont {A.}~\bibnamefont {Lewis}},\
  }\href {\doibase 10.1103/PhysRevD.67.123513} {\bibfield  {journal} {\bibinfo
  {journal} {Phys. Rev. D}\ }\textbf {\bibinfo {volume} {67}},\ \bibinfo
  {pages} {123513} (\bibinfo {year} {2003})},\ \Eprint
  {http://arxiv.org/abs/astro-ph/0212248} {arXiv:astro-ph/0212248} \BibitemShut
  {NoStop}%
\bibitem [{\citenamefont {Gordon}\ and\ \citenamefont
  {Pritchard}(2009)}]{Gordon:2009wx}%
  \BibitemOpen
  \bibfield  {author} {\bibinfo {author} {\bibfnamefont {C.}~\bibnamefont
  {Gordon}}\ and\ \bibinfo {author} {\bibfnamefont {J.~R.}\ \bibnamefont
  {Pritchard}},\ }\href {\doibase 10.1103/PhysRevD.80.063535} {\bibfield
  {journal} {\bibinfo  {journal} {Phys. Rev. D}\ }\textbf {\bibinfo {volume}
  {80}},\ \bibinfo {pages} {063535} (\bibinfo {year} {2009})},\ \Eprint
  {http://arxiv.org/abs/0907.5400} {arXiv:0907.5400 [astro-ph.CO]} \BibitemShut
  {NoStop}%
\bibitem [{\citenamefont {Holder}\ \emph {et~al.}(2010)\citenamefont {Holder},
  \citenamefont {Nollett},\ and\ \citenamefont {van Engelen}}]{Holder:2009gd}%
  \BibitemOpen
  \bibfield  {author} {\bibinfo {author} {\bibfnamefont {G.~P.}\ \bibnamefont
  {Holder}}, \bibinfo {author} {\bibfnamefont {K.~M.}\ \bibnamefont {Nollett}},
  \ and\ \bibinfo {author} {\bibfnamefont {A.}~\bibnamefont {van Engelen}},\
  }\href {\doibase 10.1088/0004-637X/716/2/907} {\bibfield  {journal} {\bibinfo
   {journal} {Astrophys. J.}\ }\textbf {\bibinfo {volume} {716}},\ \bibinfo
  {pages} {907} (\bibinfo {year} {2010})},\ \Eprint
  {http://arxiv.org/abs/0907.3919} {arXiv:0907.3919 [astro-ph.CO]} \BibitemShut
  {NoStop}%
\bibitem [{\citenamefont {Linde}\ and\ \citenamefont
  {Mukhanov}(1997)}]{Linde:1996gt}%
  \BibitemOpen
  \bibfield  {author} {\bibinfo {author} {\bibfnamefont {A.~D.}\ \bibnamefont
  {Linde}}\ and\ \bibinfo {author} {\bibfnamefont {V.~F.}\ \bibnamefont
  {Mukhanov}},\ }\href {\doibase 10.1103/PhysRevD.56.R535} {\bibfield
  {journal} {\bibinfo  {journal} {Phys. Rev. D}\ }\textbf {\bibinfo {volume}
  {56}},\ \bibinfo {pages} {R535} (\bibinfo {year} {1997})},\ \Eprint
  {http://arxiv.org/abs/astro-ph/9610219} {arXiv:astro-ph/9610219} \BibitemShut
  {NoStop}%
\bibitem [{\citenamefont {Sasaki}\ \emph {et~al.}(2006)\citenamefont {Sasaki},
  \citenamefont {Valiviita},\ and\ \citenamefont {Wands}}]{Sasaki:2006kq}%
  \BibitemOpen
  \bibfield  {author} {\bibinfo {author} {\bibfnamefont {M.}~\bibnamefont
  {Sasaki}}, \bibinfo {author} {\bibfnamefont {J.}~\bibnamefont {Valiviita}}, \
  and\ \bibinfo {author} {\bibfnamefont {D.}~\bibnamefont {Wands}},\ }\href
  {\doibase 10.1103/PhysRevD.74.103003} {\bibfield  {journal} {\bibinfo
  {journal} {Phys. Rev. D}\ }\textbf {\bibinfo {volume} {74}},\ \bibinfo
  {pages} {103003} (\bibinfo {year} {2006})},\ \Eprint
  {http://arxiv.org/abs/astro-ph/0607627} {arXiv:astro-ph/0607627} \BibitemShut
  {NoStop}%
\bibitem [{\citenamefont {Lyth}\ \emph {et~al.}(2003)\citenamefont {Lyth},
  \citenamefont {Ungarelli},\ and\ \citenamefont {Wands}}]{Lyth:2002my}%
  \BibitemOpen
  \bibfield  {author} {\bibinfo {author} {\bibfnamefont {D.~H.}\ \bibnamefont
  {Lyth}}, \bibinfo {author} {\bibfnamefont {C.}~\bibnamefont {Ungarelli}}, \
  and\ \bibinfo {author} {\bibfnamefont {D.}~\bibnamefont {Wands}},\ }\href
  {\doibase 10.1103/PhysRevD.67.023503} {\bibfield  {journal} {\bibinfo
  {journal} {Phys. Rev. D}\ }\textbf {\bibinfo {volume} {67}},\ \bibinfo
  {pages} {023503} (\bibinfo {year} {2003})},\ \Eprint
  {http://arxiv.org/abs/astro-ph/0208055} {arXiv:astro-ph/0208055} \BibitemShut
  {NoStop}%
\bibitem [{\citenamefont {Langlois}\ and\ \citenamefont
  {Riazuelo}(2000)}]{Langlois:2000ar}%
  \BibitemOpen
  \bibfield  {author} {\bibinfo {author} {\bibfnamefont {D.}~\bibnamefont
  {Langlois}}\ and\ \bibinfo {author} {\bibfnamefont {A.}~\bibnamefont
  {Riazuelo}},\ }\href {\doibase 10.1103/PhysRevD.62.043504} {\bibfield
  {journal} {\bibinfo  {journal} {Phys. Rev. D}\ }\textbf {\bibinfo {volume}
  {62}},\ \bibinfo {pages} {043504} (\bibinfo {year} {2000})},\ \Eprint
  {http://arxiv.org/abs/astro-ph/9912497} {arXiv:astro-ph/9912497} \BibitemShut
  {NoStop}%
\bibitem [{\citenamefont {He}\ \emph {et~al.}(2015)\citenamefont {He},
  \citenamefont {Grin},\ and\ \citenamefont {Hu}}]{He:2015msa}%
  \BibitemOpen
  \bibfield  {author} {\bibinfo {author} {\bibfnamefont {C.}~\bibnamefont
  {He}}, \bibinfo {author} {\bibfnamefont {D.}~\bibnamefont {Grin}}, \ and\
  \bibinfo {author} {\bibfnamefont {W.}~\bibnamefont {Hu}},\ }\href {\doibase
  10.1103/PhysRevD.92.063018} {\bibfield  {journal} {\bibinfo  {journal} {Phys.
  Rev. D}\ }\textbf {\bibinfo {volume} {92}},\ \bibinfo {pages} {063018}
  (\bibinfo {year} {2015})},\ \Eprint {http://arxiv.org/abs/1505.00639}
  {arXiv:1505.00639 [astro-ph.CO]} \BibitemShut {NoStop}%
\bibitem [{\citenamefont {De~Simone}\ and\ \citenamefont
  {Kobayashi}(2016)}]{DeSimone:2016ofp}%
  \BibitemOpen
  \bibfield  {author} {\bibinfo {author} {\bibfnamefont {A.}~\bibnamefont
  {De~Simone}}\ and\ \bibinfo {author} {\bibfnamefont {T.}~\bibnamefont
  {Kobayashi}},\ }\href {\doibase 10.1088/1475-7516/2016/08/052} {\bibfield
  {journal} {\bibinfo  {journal} {JCAP}\ }\textbf {\bibinfo {volume} {08}},\
  \bibinfo {pages} {052} (\bibinfo {year} {2016})},\ \Eprint
  {http://arxiv.org/abs/1605.00670} {arXiv:1605.00670 [hep-ph]} \BibitemShut
  {NoStop}%
\bibitem [{\citenamefont {Mu\~noz}\ \emph {et~al.}(2016)\citenamefont
  {Mu\~noz}, \citenamefont {Grin}, \citenamefont {Dai}, \citenamefont
  {Kamionkowski},\ and\ \citenamefont {Kovetz}}]{Munoz:2015fdv}%
  \BibitemOpen
  \bibfield  {author} {\bibinfo {author} {\bibfnamefont {J.~B.}\ \bibnamefont
  {Mu\~noz}}, \bibinfo {author} {\bibfnamefont {D.}~\bibnamefont {Grin}},
  \bibinfo {author} {\bibfnamefont {L.}~\bibnamefont {Dai}}, \bibinfo {author}
  {\bibfnamefont {M.}~\bibnamefont {Kamionkowski}}, \ and\ \bibinfo {author}
  {\bibfnamefont {E.~D.}\ \bibnamefont {Kovetz}},\ }\href {\doibase
  10.1103/PhysRevD.93.043008} {\bibfield  {journal} {\bibinfo  {journal} {Phys.
  Rev. D}\ }\textbf {\bibinfo {volume} {93}},\ \bibinfo {pages} {043008}
  (\bibinfo {year} {2016})},\ \Eprint {http://arxiv.org/abs/1511.04441}
  {arXiv:1511.04441 [astro-ph.CO]} \BibitemShut {NoStop}%
\bibitem [{\citenamefont {Heinrich}\ \emph {et~al.}(2016)\citenamefont
  {Heinrich}, \citenamefont {Grin},\ and\ \citenamefont
  {Hu}}]{Heinrich:2016gqe}%
  \BibitemOpen
  \bibfield  {author} {\bibinfo {author} {\bibfnamefont {C.~H.}\ \bibnamefont
  {Heinrich}}, \bibinfo {author} {\bibfnamefont {D.}~\bibnamefont {Grin}}, \
  and\ \bibinfo {author} {\bibfnamefont {W.}~\bibnamefont {Hu}},\ }\href
  {\doibase 10.1103/PhysRevD.94.043534} {\bibfield  {journal} {\bibinfo
  {journal} {Phys. Rev. D}\ }\textbf {\bibinfo {volume} {94}},\ \bibinfo
  {pages} {043534} (\bibinfo {year} {2016})},\ \Eprint
  {http://arxiv.org/abs/1605.08439} {arXiv:1605.08439 [astro-ph.CO]}
  \BibitemShut {NoStop}%
\bibitem [{\citenamefont {Smith}\ \emph {et~al.}(2017)\citenamefont {Smith},
  \citenamefont {Mu\~noz}, \citenamefont {Smith}, \citenamefont {Yee},\ and\
  \citenamefont {Grin}}]{Smith:2017ndr}%
  \BibitemOpen
  \bibfield  {author} {\bibinfo {author} {\bibfnamefont {T.~L.}\ \bibnamefont
  {Smith}}, \bibinfo {author} {\bibfnamefont {J.~B.}\ \bibnamefont {Mu\~noz}},
  \bibinfo {author} {\bibfnamefont {R.}~\bibnamefont {Smith}}, \bibinfo
  {author} {\bibfnamefont {K.}~\bibnamefont {Yee}}, \ and\ \bibinfo {author}
  {\bibfnamefont {D.}~\bibnamefont {Grin}},\ }\href {\doibase
  10.1103/PhysRevD.96.083508} {\bibfield  {journal} {\bibinfo  {journal} {Phys.
  Rev. D}\ }\textbf {\bibinfo {volume} {96}},\ \bibinfo {pages} {083508}
  (\bibinfo {year} {2017})},\ \Eprint {http://arxiv.org/abs/1704.03461}
  {arXiv:1704.03461 [astro-ph.CO]} \BibitemShut {NoStop}%
\bibitem [{\citenamefont {Valiviita}(2017)}]{Valiviita:2017fbx}%
  \BibitemOpen
  \bibfield  {author} {\bibinfo {author} {\bibfnamefont {J.}~\bibnamefont
  {Valiviita}},\ }\href {\doibase 10.1088/1475-7516/2017/04/014} {\bibfield
  {journal} {\bibinfo  {journal} {JCAP}\ }\textbf {\bibinfo {volume} {04}},\
  \bibinfo {pages} {014} (\bibinfo {year} {2017})},\ \Eprint
  {http://arxiv.org/abs/1701.07039} {arXiv:1701.07039 [astro-ph.CO]}
  \BibitemShut {NoStop}%
\bibitem [{\citenamefont {Akrami}\ \emph {et~al.}(2020)\citenamefont {Akrami}
  \emph {et~al.}}]{Planck:2018jri}%
  \BibitemOpen
  \bibfield  {author} {\bibinfo {author} {\bibfnamefont {Y.}~\bibnamefont
  {Akrami}} \emph {et~al.} (\bibinfo {collaboration} {Planck}),\ }\href
  {\doibase 10.1051/0004-6361/201833887} {\bibfield  {journal} {\bibinfo
  {journal} {Astron. Astrophys.}\ }\textbf {\bibinfo {volume} {641}},\ \bibinfo
  {pages} {A10} (\bibinfo {year} {2020})},\ \Eprint
  {http://arxiv.org/abs/1807.06211} {arXiv:1807.06211 [astro-ph.CO]}
  \BibitemShut {NoStop}%
\bibitem [{\citenamefont {Grin}\ \emph
  {et~al.}(2011{\natexlab{a}})\citenamefont {Grin}, \citenamefont {Dore},\ and\
  \citenamefont {Kamionkowski}}]{Grin:2011tf}%
  \BibitemOpen
  \bibfield  {author} {\bibinfo {author} {\bibfnamefont {D.}~\bibnamefont
  {Grin}}, \bibinfo {author} {\bibfnamefont {O.}~\bibnamefont {Dore}}, \ and\
  \bibinfo {author} {\bibfnamefont {M.}~\bibnamefont {Kamionkowski}},\ }\href
  {\doibase 10.1103/PhysRevD.84.123003} {\bibfield  {journal} {\bibinfo
  {journal} {Phys. Rev. D}\ }\textbf {\bibinfo {volume} {84}},\ \bibinfo
  {pages} {123003} (\bibinfo {year} {2011}{\natexlab{a}})},\ \Eprint
  {http://arxiv.org/abs/1107.5047} {arXiv:1107.5047 [astro-ph.CO]} \BibitemShut
  {NoStop}%
\bibitem [{\citenamefont {Grin}\ \emph
  {et~al.}(2011{\natexlab{b}})\citenamefont {Grin}, \citenamefont {Dore},\ and\
  \citenamefont {Kamionkowski}}]{Grin:2011nk}%
  \BibitemOpen
  \bibfield  {author} {\bibinfo {author} {\bibfnamefont {D.}~\bibnamefont
  {Grin}}, \bibinfo {author} {\bibfnamefont {O.}~\bibnamefont {Dore}}, \ and\
  \bibinfo {author} {\bibfnamefont {M.}~\bibnamefont {Kamionkowski}},\ }\href
  {\doibase 10.1103/PhysRevLett.107.261301} {\bibfield  {journal} {\bibinfo
  {journal} {Phys. Rev. Lett.}\ }\textbf {\bibinfo {volume} {107}},\ \bibinfo
  {pages} {261301} (\bibinfo {year} {2011}{\natexlab{b}})},\ \Eprint
  {http://arxiv.org/abs/1107.1716} {arXiv:1107.1716 [astro-ph.CO]} \BibitemShut
  {NoStop}%
\bibitem [{\citenamefont {Grin}\ \emph {et~al.}(2014)\citenamefont {Grin},
  \citenamefont {Hanson}, \citenamefont {Holder}, \citenamefont {Dor\'e},\ and\
  \citenamefont {Kamionkowski}}]{Grin:2013uya}%
  \BibitemOpen
  \bibfield  {author} {\bibinfo {author} {\bibfnamefont {D.}~\bibnamefont
  {Grin}}, \bibinfo {author} {\bibfnamefont {D.}~\bibnamefont {Hanson}},
  \bibinfo {author} {\bibfnamefont {G.~P.}\ \bibnamefont {Holder}}, \bibinfo
  {author} {\bibfnamefont {O.}~\bibnamefont {Dor\'e}}, \ and\ \bibinfo {author}
  {\bibfnamefont {M.}~\bibnamefont {Kamionkowski}},\ }\href {\doibase
  10.1103/PhysRevD.89.023006} {\bibfield  {journal} {\bibinfo  {journal} {Phys.
  Rev. D}\ }\textbf {\bibinfo {volume} {89}},\ \bibinfo {pages} {023006}
  (\bibinfo {year} {2014})},\ \Eprint {http://arxiv.org/abs/1306.4319}
  {arXiv:1306.4319 [astro-ph.CO]} \BibitemShut {NoStop}%
\bibitem [{\citenamefont {Haga}\ \emph {et~al.}(2018)\citenamefont {Haga},
  \citenamefont {Inomata}, \citenamefont {Ota},\ and\ \citenamefont
  {Ravenni}}]{Haga:2018pdl}%
  \BibitemOpen
  \bibfield  {author} {\bibinfo {author} {\bibfnamefont {T.}~\bibnamefont
  {Haga}}, \bibinfo {author} {\bibfnamefont {K.}~\bibnamefont {Inomata}},
  \bibinfo {author} {\bibfnamefont {A.}~\bibnamefont {Ota}}, \ and\ \bibinfo
  {author} {\bibfnamefont {A.}~\bibnamefont {Ravenni}},\ }\href {\doibase
  10.1088/1475-7516/2018/08/036} {\bibfield  {journal} {\bibinfo  {journal}
  {JCAP}\ }\textbf {\bibinfo {volume} {08}},\ \bibinfo {pages} {036} (\bibinfo
  {year} {2018})},\ \Eprint {http://arxiv.org/abs/1805.08773} {arXiv:1805.08773
  [astro-ph.CO]} \BibitemShut {NoStop}%
\bibitem [{\citenamefont {Chluba}\ and\ \citenamefont
  {Grin}(2013)}]{Chluba:2013dna}%
  \BibitemOpen
  \bibfield  {author} {\bibinfo {author} {\bibfnamefont {J.}~\bibnamefont
  {Chluba}}\ and\ \bibinfo {author} {\bibfnamefont {D.}~\bibnamefont {Grin}},\
  }\href {\doibase 10.1093/mnras/stt1129} {\bibfield  {journal} {\bibinfo
  {journal} {Mon. Not. Roy. Astron. Soc.}\ }\textbf {\bibinfo {volume} {434}},\
  \bibinfo {pages} {1619} (\bibinfo {year} {2013})},\ \Eprint
  {http://arxiv.org/abs/1304.4596} {arXiv:1304.4596 [astro-ph.CO]} \BibitemShut
  {NoStop}%
\bibitem [{\citenamefont {Lee}\ and\ \citenamefont
  {Ali-Ha\"\i{}moud}(2021)}]{Lee:2021bmn}%
  \BibitemOpen
  \bibfield  {author} {\bibinfo {author} {\bibfnamefont {N.}~\bibnamefont
  {Lee}}\ and\ \bibinfo {author} {\bibfnamefont {Y.}~\bibnamefont
  {Ali-Ha\"\i{}moud}},\ }\href {\doibase 10.1103/PhysRevD.104.103509}
  {\bibfield  {journal} {\bibinfo  {journal} {Phys. Rev. D}\ }\textbf {\bibinfo
  {volume} {104}},\ \bibinfo {pages} {103509} (\bibinfo {year} {2021})},\
  \Eprint {http://arxiv.org/abs/2108.07798} {arXiv:2108.07798 [astro-ph.CO]}
  \BibitemShut {NoStop}%
\bibitem [{\citenamefont {Soumagnac}\ \emph {et~al.}(2016)\citenamefont
  {Soumagnac}, \citenamefont {Barkana}, \citenamefont {Sabiu}, \citenamefont
  {Loeb}, \citenamefont {Ross}, \citenamefont {Abdalla}, \citenamefont
  {Balan},\ and\ \citenamefont {Lahav}}]{Soumagnac:2016bjk}%
  \BibitemOpen
  \bibfield  {author} {\bibinfo {author} {\bibfnamefont {M.~T.}\ \bibnamefont
  {Soumagnac}}, \bibinfo {author} {\bibfnamefont {R.}~\bibnamefont {Barkana}},
  \bibinfo {author} {\bibfnamefont {C.~G.}\ \bibnamefont {Sabiu}}, \bibinfo
  {author} {\bibfnamefont {A.}~\bibnamefont {Loeb}}, \bibinfo {author}
  {\bibfnamefont {A.~J.}\ \bibnamefont {Ross}}, \bibinfo {author}
  {\bibfnamefont {F.~B.}\ \bibnamefont {Abdalla}}, \bibinfo {author}
  {\bibfnamefont {S.~T.}\ \bibnamefont {Balan}}, \ and\ \bibinfo {author}
  {\bibfnamefont {O.}~\bibnamefont {Lahav}},\ }\href {\doibase
  10.1103/PhysRevLett.116.201302} {\bibfield  {journal} {\bibinfo  {journal}
  {Phys. Rev. Lett.}\ }\textbf {\bibinfo {volume} {116}},\ \bibinfo {pages}
  {201302} (\bibinfo {year} {2016})},\ \Eprint
  {http://arxiv.org/abs/1602.01839} {arXiv:1602.01839 [astro-ph.CO]}
  \BibitemShut {NoStop}%
\bibitem [{\citenamefont {Soumagnac}\ \emph {et~al.}(2018)\citenamefont
  {Soumagnac}, \citenamefont {Sabiu}, \citenamefont {Barkana},\ and\
  \citenamefont {Yoo}}]{Soumagnac:2018atx}%
  \BibitemOpen
  \bibfield  {author} {\bibinfo {author} {\bibfnamefont {M.~T.}\ \bibnamefont
  {Soumagnac}}, \bibinfo {author} {\bibfnamefont {C.~G.}\ \bibnamefont
  {Sabiu}}, \bibinfo {author} {\bibfnamefont {R.}~\bibnamefont {Barkana}}, \
  and\ \bibinfo {author} {\bibfnamefont {J.}~\bibnamefont {Yoo}},\ }\href
  {\doibase 10.1093/mnras/stz240} {\  (\bibinfo {year} {2018}),\
  10.1093/mnras/stz240},\ \Eprint {http://arxiv.org/abs/1802.10368}
  {arXiv:1802.10368 [astro-ph.CO]} \BibitemShut {NoStop}%
\bibitem [{\citenamefont {Heinrich}\ and\ \citenamefont
  {Schmittfull}(2019)}]{Heinrich:2019sxl}%
  \BibitemOpen
  \bibfield  {author} {\bibinfo {author} {\bibfnamefont {C.}~\bibnamefont
  {Heinrich}}\ and\ \bibinfo {author} {\bibfnamefont {M.}~\bibnamefont
  {Schmittfull}},\ }\href {\doibase 10.1103/PhysRevD.100.063503} {\bibfield
  {journal} {\bibinfo  {journal} {Phys. Rev. D}\ }\textbf {\bibinfo {volume}
  {100}},\ \bibinfo {pages} {063503} (\bibinfo {year} {2019})},\ \Eprint
  {http://arxiv.org/abs/1904.00024} {arXiv:1904.00024 [astro-ph.CO]}
  \BibitemShut {NoStop}%
\bibitem [{\citenamefont {Mu\~noz}(2019{\natexlab{a}})}]{Munoz:2019rhi}%
  \BibitemOpen
  \bibfield  {author} {\bibinfo {author} {\bibfnamefont {J.~B.}\ \bibnamefont
  {Mu\~noz}},\ }\href {\doibase 10.1103/PhysRevD.100.063538} {\bibfield
  {journal} {\bibinfo  {journal} {Phys. Rev. D}\ }\textbf {\bibinfo {volume}
  {100}},\ \bibinfo {pages} {063538} (\bibinfo {year} {2019}{\natexlab{a}})},\
  \Eprint {http://arxiv.org/abs/1904.07881} {arXiv:1904.07881 [astro-ph.CO]}
  \BibitemShut {NoStop}%
\bibitem [{\citenamefont {Mu\~noz}(2019{\natexlab{b}})}]{Munoz:2019fkt}%
  \BibitemOpen
  \bibfield  {author} {\bibinfo {author} {\bibfnamefont {J.~B.}\ \bibnamefont
  {Mu\~noz}},\ }\href {\doibase 10.1103/PhysRevLett.123.131301} {\bibfield
  {journal} {\bibinfo  {journal} {Phys. Rev. Lett.}\ }\textbf {\bibinfo
  {volume} {123}},\ \bibinfo {pages} {131301} (\bibinfo {year}
  {2019}{\natexlab{b}})},\ \Eprint {http://arxiv.org/abs/1904.07868}
  {arXiv:1904.07868 [astro-ph.CO]} \BibitemShut {NoStop}%
\bibitem [{\citenamefont {Hotinli}\ \emph {et~al.}(2021)\citenamefont
  {Hotinli}, \citenamefont {Binnie}, \citenamefont {Mu\~noz}, \citenamefont
  {Dinda},\ and\ \citenamefont {Kamionkowski}}]{Hotinli:2021xln}%
  \BibitemOpen
  \bibfield  {author} {\bibinfo {author} {\bibfnamefont {S.~C.}\ \bibnamefont
  {Hotinli}}, \bibinfo {author} {\bibfnamefont {T.}~\bibnamefont {Binnie}},
  \bibinfo {author} {\bibfnamefont {J.~B.}\ \bibnamefont {Mu\~noz}}, \bibinfo
  {author} {\bibfnamefont {B.~R.}\ \bibnamefont {Dinda}}, \ and\ \bibinfo
  {author} {\bibfnamefont {M.}~\bibnamefont {Kamionkowski}},\ }\href {\doibase
  10.1103/PhysRevD.104.063536} {\bibfield  {journal} {\bibinfo  {journal}
  {Phys. Rev. D}\ }\textbf {\bibinfo {volume} {104}},\ \bibinfo {pages}
  {063536} (\bibinfo {year} {2021})},\ \Eprint
  {http://arxiv.org/abs/2106.11979} {arXiv:2106.11979 [astro-ph.CO]}
  \BibitemShut {NoStop}%
\bibitem [{\citenamefont {Barreira}\ \emph
  {et~al.}(2020{\natexlab{a}})\citenamefont {Barreira}, \citenamefont {Cabass},
  \citenamefont {Nelson},\ and\ \citenamefont {Schmidt}}]{Barreira:2019qdl}%
  \BibitemOpen
  \bibfield  {author} {\bibinfo {author} {\bibfnamefont {A.}~\bibnamefont
  {Barreira}}, \bibinfo {author} {\bibfnamefont {G.}~\bibnamefont {Cabass}},
  \bibinfo {author} {\bibfnamefont {D.}~\bibnamefont {Nelson}}, \ and\ \bibinfo
  {author} {\bibfnamefont {F.}~\bibnamefont {Schmidt}},\ }\href {\doibase
  10.1088/1475-7516/2020/02/005} {\bibfield  {journal} {\bibinfo  {journal}
  {JCAP}\ }\textbf {\bibinfo {volume} {02}},\ \bibinfo {pages} {005} (\bibinfo
  {year} {2020}{\natexlab{a}})},\ \Eprint {http://arxiv.org/abs/1907.04317}
  {arXiv:1907.04317 [astro-ph.CO]} \BibitemShut {NoStop}%
\bibitem [{\citenamefont {Barreira}\ \emph
  {et~al.}(2020{\natexlab{b}})\citenamefont {Barreira}, \citenamefont {Cabass},
  \citenamefont {Lozanov},\ and\ \citenamefont {Schmidt}}]{Barreira:2020lva}%
  \BibitemOpen
  \bibfield  {author} {\bibinfo {author} {\bibfnamefont {A.}~\bibnamefont
  {Barreira}}, \bibinfo {author} {\bibfnamefont {G.}~\bibnamefont {Cabass}},
  \bibinfo {author} {\bibfnamefont {K.~D.}\ \bibnamefont {Lozanov}}, \ and\
  \bibinfo {author} {\bibfnamefont {F.}~\bibnamefont {Schmidt}},\ }\href
  {\doibase 10.1088/1475-7516/2020/07/049} {\bibfield  {journal} {\bibinfo
  {journal} {JCAP}\ }\textbf {\bibinfo {volume} {07}},\ \bibinfo {pages} {049}
  (\bibinfo {year} {2020}{\natexlab{b}})},\ \Eprint
  {http://arxiv.org/abs/2002.12931} {arXiv:2002.12931 [astro-ph.CO]}
  \BibitemShut {NoStop}%
\bibitem [{\citenamefont {Hotinli}\ \emph {et~al.}(2019)\citenamefont
  {Hotinli}, \citenamefont {Mertens}, \citenamefont {Johnson},\ and\
  \citenamefont {Kamionkowski}}]{Hotinli:2019wdp}%
  \BibitemOpen
  \bibfield  {author} {\bibinfo {author} {\bibfnamefont {S.~C.}\ \bibnamefont
  {Hotinli}}, \bibinfo {author} {\bibfnamefont {J.~B.}\ \bibnamefont
  {Mertens}}, \bibinfo {author} {\bibfnamefont {M.~C.}\ \bibnamefont
  {Johnson}}, \ and\ \bibinfo {author} {\bibfnamefont {M.}~\bibnamefont
  {Kamionkowski}},\ }\href {\doibase 10.1103/PhysRevD.100.103528} {\bibfield
  {journal} {\bibinfo  {journal} {Phys. Rev. D}\ }\textbf {\bibinfo {volume}
  {100}},\ \bibinfo {pages} {103528} (\bibinfo {year} {2019})},\ \Eprint
  {http://arxiv.org/abs/1908.08953} {arXiv:1908.08953 [astro-ph.CO]}
  \BibitemShut {NoStop}%
\bibitem [{\citenamefont {Sato-Polito}\ \emph {et~al.}(2021)\citenamefont
  {Sato-Polito}, \citenamefont {Bernal}, \citenamefont {Boddy},\ and\
  \citenamefont {Kamionkowski}}]{Sato-Polito:2020cil}%
  \BibitemOpen
  \bibfield  {author} {\bibinfo {author} {\bibfnamefont {G.}~\bibnamefont
  {Sato-Polito}}, \bibinfo {author} {\bibfnamefont {J.~L.}\ \bibnamefont
  {Bernal}}, \bibinfo {author} {\bibfnamefont {K.~K.}\ \bibnamefont {Boddy}}, \
  and\ \bibinfo {author} {\bibfnamefont {M.}~\bibnamefont {Kamionkowski}},\
  }\href {\doibase 10.1103/PhysRevD.103.083519} {\bibfield  {journal} {\bibinfo
   {journal} {Phys. Rev. D}\ }\textbf {\bibinfo {volume} {103}},\ \bibinfo
  {pages} {083519} (\bibinfo {year} {2021})},\ \Eprint
  {http://arxiv.org/abs/2011.08193} {arXiv:2011.08193 [astro-ph.CO]}
  \BibitemShut {NoStop}%
\bibitem [{\citenamefont {Kumar}\ \emph {et~al.}(2023)\citenamefont {Kumar},
  \citenamefont {Hotinli},\ and\ \citenamefont {Kamionkowski}}]{Kumar:2022bly}%
  \BibitemOpen
  \bibfield  {author} {\bibinfo {author} {\bibfnamefont {N.~A.}\ \bibnamefont
  {Kumar}}, \bibinfo {author} {\bibfnamefont {S.~C.}\ \bibnamefont {Hotinli}},
  \ and\ \bibinfo {author} {\bibfnamefont {M.}~\bibnamefont {Kamionkowski}},\
  }\href {\doibase 10.1103/PhysRevD.107.043504} {\bibfield  {journal} {\bibinfo
   {journal} {Phys. Rev. D}\ }\textbf {\bibinfo {volume} {107}},\ \bibinfo
  {pages} {043504} (\bibinfo {year} {2023})},\ \Eprint
  {http://arxiv.org/abs/2208.02829} {arXiv:2208.02829 [astro-ph.CO]}
  \BibitemShut {NoStop}%
\bibitem [{\citenamefont {Naoz}\ and\ \citenamefont
  {Narayan}(2013)}]{Naoz:2013wla}%
  \BibitemOpen
  \bibfield  {author} {\bibinfo {author} {\bibfnamefont {S.}~\bibnamefont
  {Naoz}}\ and\ \bibinfo {author} {\bibfnamefont {R.}~\bibnamefont {Narayan}},\
  }\href {\doibase 10.1103/PhysRevLett.111.051303} {\bibfield  {journal}
  {\bibinfo  {journal} {Phys. Rev. Lett.}\ }\textbf {\bibinfo {volume} {111}},\
  \bibinfo {pages} {051303} (\bibinfo {year} {2013})},\ \Eprint
  {http://arxiv.org/abs/1304.5792} {arXiv:1304.5792 [astro-ph.CO]} \BibitemShut
  {NoStop}%
\bibitem [{\citenamefont {Venumadhav}\ \emph {et~al.}(2017)\citenamefont
  {Venumadhav}, \citenamefont {Oklopcic}, \citenamefont {Gluscevic},
  \citenamefont {Mishra},\ and\ \citenamefont {Hirata}}]{Venumadhav:2014tqa}%
  \BibitemOpen
  \bibfield  {author} {\bibinfo {author} {\bibfnamefont {T.}~\bibnamefont
  {Venumadhav}}, \bibinfo {author} {\bibfnamefont {A.}~\bibnamefont
  {Oklopcic}}, \bibinfo {author} {\bibfnamefont {V.}~\bibnamefont {Gluscevic}},
  \bibinfo {author} {\bibfnamefont {A.}~\bibnamefont {Mishra}}, \ and\ \bibinfo
  {author} {\bibfnamefont {C.~M.}\ \bibnamefont {Hirata}},\ }\href {\doibase
  10.1103/PhysRevD.95.083010} {\bibfield  {journal} {\bibinfo  {journal} {Phys.
  Rev. D}\ }\textbf {\bibinfo {volume} {95}},\ \bibinfo {pages} {083010}
  (\bibinfo {year} {2017})},\ \Eprint {http://arxiv.org/abs/1410.2250}
  {arXiv:1410.2250 [astro-ph.CO]} \BibitemShut {NoStop}%
\bibitem [{\citenamefont {Gluscevic}\ \emph {et~al.}(2017)\citenamefont
  {Gluscevic}, \citenamefont {Venumadhav}, \citenamefont {Fang}, \citenamefont
  {Hirata}, \citenamefont {Oklopcic},\ and\ \citenamefont
  {Mishra}}]{Gluscevic:2016gns}%
  \BibitemOpen
  \bibfield  {author} {\bibinfo {author} {\bibfnamefont {V.}~\bibnamefont
  {Gluscevic}}, \bibinfo {author} {\bibfnamefont {T.}~\bibnamefont
  {Venumadhav}}, \bibinfo {author} {\bibfnamefont {X.}~\bibnamefont {Fang}},
  \bibinfo {author} {\bibfnamefont {C.}~\bibnamefont {Hirata}}, \bibinfo
  {author} {\bibfnamefont {A.}~\bibnamefont {Oklopcic}}, \ and\ \bibinfo
  {author} {\bibfnamefont {A.}~\bibnamefont {Mishra}},\ }\href {\doibase
  10.1103/PhysRevD.95.083011} {\bibfield  {journal} {\bibinfo  {journal} {Phys.
  Rev. D}\ }\textbf {\bibinfo {volume} {95}},\ \bibinfo {pages} {083011}
  (\bibinfo {year} {2017})},\ \Eprint {http://arxiv.org/abs/1604.06327}
  {arXiv:1604.06327 [astro-ph.CO]} \BibitemShut {NoStop}%
\bibitem [{\citenamefont {Aghanim}\ \emph {et~al.}(2020)\citenamefont {Aghanim}
  \emph {et~al.}}]{Planck:2018vyg}%
  \BibitemOpen
  \bibfield  {author} {\bibinfo {author} {\bibfnamefont {N.}~\bibnamefont
  {Aghanim}} \emph {et~al.} (\bibinfo {collaboration} {Planck}),\ }\href
  {\doibase 10.1051/0004-6361/201833910} {\bibfield  {journal} {\bibinfo
  {journal} {Astron. Astrophys.}\ }\textbf {\bibinfo {volume} {641}},\ \bibinfo
  {pages} {A6} (\bibinfo {year} {2020})},\ \bibinfo {note} {[Erratum:
  Astron.Astrophys. 652, C4 (2021)]},\ \Eprint
  {http://arxiv.org/abs/1807.06209} {arXiv:1807.06209 [astro-ph.CO]}
  \BibitemShut {NoStop}%
\bibitem [{\citenamefont {Cyburt}\ \emph {et~al.}(2016)\citenamefont {Cyburt},
  \citenamefont {Fields}, \citenamefont {Olive},\ and\ \citenamefont
  {Yeh}}]{Cyburt:2015mya}%
  \BibitemOpen
  \bibfield  {author} {\bibinfo {author} {\bibfnamefont {R.~H.}\ \bibnamefont
  {Cyburt}}, \bibinfo {author} {\bibfnamefont {B.~D.}\ \bibnamefont {Fields}},
  \bibinfo {author} {\bibfnamefont {K.~A.}\ \bibnamefont {Olive}}, \ and\
  \bibinfo {author} {\bibfnamefont {T.-H.}\ \bibnamefont {Yeh}},\ }\href
  {\doibase 10.1103/RevModPhys.88.015004} {\bibfield  {journal} {\bibinfo
  {journal} {Rev. Mod. Phys.}\ }\textbf {\bibinfo {volume} {88}},\ \bibinfo
  {pages} {015004} (\bibinfo {year} {2016})},\ \Eprint
  {http://arxiv.org/abs/1505.01076} {arXiv:1505.01076 [astro-ph.CO]}
  \BibitemShut {NoStop}%
\bibitem [{\citenamefont {Cooke}\ \emph {et~al.}(2018)\citenamefont {Cooke},
  \citenamefont {Pettini},\ and\ \citenamefont {Steidel}}]{Cooke:2017cwo}%
  \BibitemOpen
  \bibfield  {author} {\bibinfo {author} {\bibfnamefont {R.~J.}\ \bibnamefont
  {Cooke}}, \bibinfo {author} {\bibfnamefont {M.}~\bibnamefont {Pettini}}, \
  and\ \bibinfo {author} {\bibfnamefont {C.~C.}\ \bibnamefont {Steidel}},\
  }\href {\doibase 10.3847/1538-4357/aaab53} {\bibfield  {journal} {\bibinfo
  {journal} {Astrophys. J.}\ }\textbf {\bibinfo {volume} {855}},\ \bibinfo
  {pages} {102} (\bibinfo {year} {2018})},\ \Eprint
  {http://arxiv.org/abs/1710.11129} {arXiv:1710.11129 [astro-ph.CO]}
  \BibitemShut {NoStop}%
\bibitem [{\citenamefont {Ali-Haimoud}\ and\ \citenamefont
  {Hirata}(2011)}]{Ali-Haimoud:2010hou}%
  \BibitemOpen
  \bibfield  {author} {\bibinfo {author} {\bibfnamefont {Y.}~\bibnamefont
  {Ali-Haimoud}}\ and\ \bibinfo {author} {\bibfnamefont {C.~M.}\ \bibnamefont
  {Hirata}},\ }\href {\doibase 10.1103/PhysRevD.83.043513} {\bibfield
  {journal} {\bibinfo  {journal} {Phys. Rev. D}\ }\textbf {\bibinfo {volume}
  {83}},\ \bibinfo {pages} {043513} (\bibinfo {year} {2011})},\ \Eprint
  {http://arxiv.org/abs/1011.3758} {arXiv:1011.3758 [astro-ph.CO]} \BibitemShut
  {NoStop}%
\bibitem [{\citenamefont {Lee}\ and\ \citenamefont
  {Ali-Ha\"\i{}moud}(2020)}]{Lee:2020obi}%
  \BibitemOpen
  \bibfield  {author} {\bibinfo {author} {\bibfnamefont {N.}~\bibnamefont
  {Lee}}\ and\ \bibinfo {author} {\bibfnamefont {Y.}~\bibnamefont
  {Ali-Ha\"\i{}moud}},\ }\href {\doibase 10.1103/PhysRevD.102.083517}
  {\bibfield  {journal} {\bibinfo  {journal} {Phys. Rev. D}\ }\textbf {\bibinfo
  {volume} {102}},\ \bibinfo {pages} {083517} (\bibinfo {year} {2020})},\
  \Eprint {http://arxiv.org/abs/2007.14114} {arXiv:2007.14114 [astro-ph.CO]}
  \BibitemShut {NoStop}%
\bibitem [{\citenamefont {Lesgourgues}(2011)}]{Lesgourgues:2011re}%
  \BibitemOpen
  \bibfield  {author} {\bibinfo {author} {\bibfnamefont {J.}~\bibnamefont
  {Lesgourgues}},\ }\href@noop {} {\  (\bibinfo {year} {2011})},\ \Eprint
  {http://arxiv.org/abs/1104.2932} {arXiv:1104.2932 [astro-ph.IM]} \BibitemShut
  {NoStop}%
\bibitem [{\citenamefont {Naoz}\ \emph {et~al.}(2011)\citenamefont {Naoz},
  \citenamefont {Yoshida},\ and\ \citenamefont {Barkana}}]{Naoz:2010hg}%
  \BibitemOpen
  \bibfield  {author} {\bibinfo {author} {\bibfnamefont {S.}~\bibnamefont
  {Naoz}}, \bibinfo {author} {\bibfnamefont {N.}~\bibnamefont {Yoshida}}, \
  and\ \bibinfo {author} {\bibfnamefont {R.}~\bibnamefont {Barkana}},\ }\href
  {\doibase 10.1111/j.1365-2966.2011.19025.x} {\bibfield  {journal} {\bibinfo
  {journal} {Mon. Not. Roy. Astron. Soc.}\ }\textbf {\bibinfo {volume} {416}},\
  \bibinfo {pages} {232} (\bibinfo {year} {2011})},\ \Eprint
  {http://arxiv.org/abs/1009.0945} {arXiv:1009.0945 [astro-ph.CO]} \BibitemShut
  {NoStop}%
\bibitem [{\citenamefont {Naoz}\ and\ \citenamefont
  {Barkana}(2005)}]{Naoz:2005pd}%
  \BibitemOpen
  \bibfield  {author} {\bibinfo {author} {\bibfnamefont {S.}~\bibnamefont
  {Naoz}}\ and\ \bibinfo {author} {\bibfnamefont {R.}~\bibnamefont {Barkana}},\
  }\href {\doibase 10.1111/j.1365-2966.2005.09385.x} {\bibfield  {journal}
  {\bibinfo  {journal} {Mon. Not. Roy. Astron. Soc.}\ }\textbf {\bibinfo
  {volume} {362}},\ \bibinfo {pages} {1047} (\bibinfo {year} {2005})},\ \Eprint
  {http://arxiv.org/abs/astro-ph/0503196} {arXiv:astro-ph/0503196} \BibitemShut
  {NoStop}%
\bibitem [{\citenamefont {{Kulsrud}}(2004)}]{2004ppa..book.....K}%
  \BibitemOpen
  \bibfield  {author} {\bibinfo {author} {\bibfnamefont {R.~M.}\ \bibnamefont
  {{Kulsrud}}},\ }\href@noop {} {\emph {\bibinfo {title} {{Plasma Physics for
  Astrophysics}}}}\ (\bibinfo {year} {2004})\BibitemShut {NoStop}%
\bibitem [{\citenamefont {Papanikolaou}\ and\ \citenamefont
  {Gourgouliatos}(2023)}]{Papanikolaou:2023nkx}%
  \BibitemOpen
  \bibfield  {author} {\bibinfo {author} {\bibfnamefont {T.}~\bibnamefont
  {Papanikolaou}}\ and\ \bibinfo {author} {\bibfnamefont {K.~N.}\ \bibnamefont
  {Gourgouliatos}},\ }\href {\doibase 10.1103/PhysRevD.107.103532} {\bibfield
  {journal} {\bibinfo  {journal} {Phys. Rev. D}\ }\textbf {\bibinfo {volume}
  {107}},\ \bibinfo {pages} {103532} (\bibinfo {year} {2023})},\ \Eprint
  {http://arxiv.org/abs/2301.10045} {arXiv:2301.10045 [astro-ph.CO]}
  \BibitemShut {NoStop}%
\bibitem [{\citenamefont {Bardeen}\ \emph {et~al.}(1986)\citenamefont
  {Bardeen}, \citenamefont {Bond}, \citenamefont {Kaiser},\ and\ \citenamefont
  {Szalay}}]{Bardeen:1985tr}%
  \BibitemOpen
  \bibfield  {author} {\bibinfo {author} {\bibfnamefont {J.~M.}\ \bibnamefont
  {Bardeen}}, \bibinfo {author} {\bibfnamefont {J.~R.}\ \bibnamefont {Bond}},
  \bibinfo {author} {\bibfnamefont {N.}~\bibnamefont {Kaiser}}, \ and\ \bibinfo
  {author} {\bibfnamefont {A.~S.}\ \bibnamefont {Szalay}},\ }\href {\doibase
  10.1086/164143} {\bibfield  {journal} {\bibinfo  {journal} {Astrophys. J.}\
  }\textbf {\bibinfo {volume} {304}},\ \bibinfo {pages} {15} (\bibinfo {year}
  {1986})}\BibitemShut {NoStop}%
\bibitem [{\citenamefont {Sugiyama}(1995)}]{Sugiyama:1994ed}%
  \BibitemOpen
  \bibfield  {author} {\bibinfo {author} {\bibfnamefont {N.}~\bibnamefont
  {Sugiyama}},\ }\href {\doibase 10.1086/192220} {\bibfield  {journal}
  {\bibinfo  {journal} {Astrophys. J. Suppl.}\ }\textbf {\bibinfo {volume}
  {100}},\ \bibinfo {pages} {281} (\bibinfo {year} {1995})},\ \Eprint
  {http://arxiv.org/abs/astro-ph/9412025} {arXiv:astro-ph/9412025} \BibitemShut
  {NoStop}%
\bibitem [{\citenamefont {Amaral}\ \emph {et~al.}(2021)\citenamefont {Amaral},
  \citenamefont {Vernstrom},\ and\ \citenamefont {Gaensler}}]{Amaral:2021mly}%
  \BibitemOpen
  \bibfield  {author} {\bibinfo {author} {\bibfnamefont {A.~D.}\ \bibnamefont
  {Amaral}}, \bibinfo {author} {\bibfnamefont {T.}~\bibnamefont {Vernstrom}}, \
  and\ \bibinfo {author} {\bibfnamefont {B.~M.}\ \bibnamefont {Gaensler}},\
  }\href {\doibase 10.1093/mnras/stab564} {\bibfield  {journal} {\bibinfo
  {journal} {Mon. Not. Roy. Astron. Soc.}\ }\textbf {\bibinfo {volume} {503}},\
  \bibinfo {pages} {2913} (\bibinfo {year} {2021})},\ \Eprint
  {http://arxiv.org/abs/2102.11312} {arXiv:2102.11312 [astro-ph.CO]}
  \BibitemShut {NoStop}%
\bibitem [{\citenamefont {Pandey}\ and\ \citenamefont
  {Sethi}(2013)}]{Pandey:2012ss}%
  \BibitemOpen
  \bibfield  {author} {\bibinfo {author} {\bibfnamefont {K.~L.}\ \bibnamefont
  {Pandey}}\ and\ \bibinfo {author} {\bibfnamefont {S.~K.}\ \bibnamefont
  {Sethi}},\ }\href {\doibase 10.1088/0004-637X/762/1/15} {\bibfield  {journal}
  {\bibinfo  {journal} {Astrophys. J.}\ }\textbf {\bibinfo {volume} {762}},\
  \bibinfo {pages} {15} (\bibinfo {year} {2013})},\ \Eprint
  {http://arxiv.org/abs/1210.3298} {arXiv:1210.3298 [astro-ph.CO]} \BibitemShut
  {NoStop}%
\bibitem [{\citenamefont {Chongchitnan}\ and\ \citenamefont
  {Meiksin}(2014)}]{Chongchitnan:2013vpa}%
  \BibitemOpen
  \bibfield  {author} {\bibinfo {author} {\bibfnamefont {S.}~\bibnamefont
  {Chongchitnan}}\ and\ \bibinfo {author} {\bibfnamefont {A.}~\bibnamefont
  {Meiksin}},\ }\href {\doibase 10.1093/mnras/stt2169} {\bibfield  {journal}
  {\bibinfo  {journal} {Mon. Not. Roy. Astron. Soc.}\ }\textbf {\bibinfo
  {volume} {437}},\ \bibinfo {pages} {3639} (\bibinfo {year} {2014})},\ \Eprint
  {http://arxiv.org/abs/1311.1504} {arXiv:1311.1504 [astro-ph.CO]} \BibitemShut
  {NoStop}%
\bibitem [{\citenamefont {Alves~Batista}\ and\ \citenamefont
  {Saveliev}(2020)}]{AlvesBatista:2020oio}%
  \BibitemOpen
  \bibfield  {author} {\bibinfo {author} {\bibfnamefont {R.}~\bibnamefont
  {Alves~Batista}}\ and\ \bibinfo {author} {\bibfnamefont {A.}~\bibnamefont
  {Saveliev}},\ }\href {\doibase 10.3847/2041-8213/abb816} {\bibfield
  {journal} {\bibinfo  {journal} {Astrophys. J. Lett.}\ }\textbf {\bibinfo
  {volume} {902}},\ \bibinfo {pages} {L11} (\bibinfo {year} {2020})},\ \Eprint
  {http://arxiv.org/abs/2009.12161} {arXiv:2009.12161 [astro-ph.HE]}
  \BibitemShut {NoStop}%
\bibitem [{\citenamefont {Ackermann}\ \emph {et~al.}(2018)\citenamefont
  {Ackermann} \emph {et~al.}}]{Fermi-LAT:2018jdy}%
  \BibitemOpen
  \bibfield  {author} {\bibinfo {author} {\bibfnamefont {M.}~\bibnamefont
  {Ackermann}} \emph {et~al.} (\bibinfo {collaboration} {Fermi-LAT}),\ }\href
  {\doibase 10.3847/1538-4365/aacdf7} {\bibfield  {journal} {\bibinfo
  {journal} {Astrophys. J. Suppl.}\ }\textbf {\bibinfo {volume} {237}},\
  \bibinfo {pages} {32} (\bibinfo {year} {2018})},\ \Eprint
  {http://arxiv.org/abs/1804.08035} {arXiv:1804.08035 [astro-ph.HE]}
  \BibitemShut {NoStop}%
\bibitem [{\citenamefont {Acciari}\ \emph {et~al.}(2022)\citenamefont {Acciari}
  \emph {et~al.}}]{MAGIC:2022piy}%
  \BibitemOpen
  \bibfield  {author} {\bibinfo {author} {\bibfnamefont {V.~A.}\ \bibnamefont
  {Acciari}} \emph {et~al.} (\bibinfo {collaboration} {MAGIC}),\ }\href@noop {}
  {\  (\bibinfo {year} {2022})},\ \Eprint {http://arxiv.org/abs/2210.03321}
  {arXiv:2210.03321 [astro-ph.HE]} \BibitemShut {NoStop}%
\bibitem [{\citenamefont {Widrow}(2002)}]{Widrow:2002ud}%
  \BibitemOpen
  \bibfield  {author} {\bibinfo {author} {\bibfnamefont {L.~M.}\ \bibnamefont
  {Widrow}},\ }\href {\doibase 10.1103/RevModPhys.74.775} {\bibfield  {journal}
  {\bibinfo  {journal} {Rev. Mod. Phys.}\ }\textbf {\bibinfo {volume} {74}},\
  \bibinfo {pages} {775} (\bibinfo {year} {2002})},\ \Eprint
  {http://arxiv.org/abs/astro-ph/0207240} {arXiv:astro-ph/0207240} \BibitemShut
  {NoStop}%
\bibitem [{\citenamefont {Davis}\ \emph {et~al.}(1999)\citenamefont {Davis},
  \citenamefont {Lilley},\ and\ \citenamefont {Tornkvist}}]{Davis:1999bt}%
  \BibitemOpen
  \bibfield  {author} {\bibinfo {author} {\bibfnamefont {A.-C.}\ \bibnamefont
  {Davis}}, \bibinfo {author} {\bibfnamefont {M.}~\bibnamefont {Lilley}}, \
  and\ \bibinfo {author} {\bibfnamefont {O.}~\bibnamefont {Tornkvist}},\ }\href
  {\doibase 10.1103/PhysRevD.60.021301} {\bibfield  {journal} {\bibinfo
  {journal} {Phys. Rev. D}\ }\textbf {\bibinfo {volume} {60}},\ \bibinfo
  {pages} {021301} (\bibinfo {year} {1999})},\ \Eprint
  {http://arxiv.org/abs/astro-ph/9904022} {arXiv:astro-ph/9904022} \BibitemShut
  {NoStop}%
\end{thebibliography}%

\end{document}